%% file: main.tex
\documentclass[conference]{IEEEtran}
\usepackage{tabularx}
\usepackage{mathrsfs}
\usepackage{tikz}
\usepackage{amsmath}
\usepackage{comment}
\let\labelindent\relax
\usepackage{enumitem}
\usepackage{algorithm}
\usepackage{algpseudocode}
\usepackage{float}

\usepackage{caption}
\usepackage{subcaption}

\usepackage{filecontents}
\usepackage{siunitx}
\usepackage{varwidth}
\usepackage{amsmath}
\usepackage{multirow}
\usepackage{url}
\usepackage{framed}
\usepackage{mdframed}
\usepackage{mathtools, nccmath}
\usepackage{balance}
\usepackage{threeparttable}
\usepackage{amsfonts}
\usepackage{mathrsfs}
\usepackage{tikz}
\usepackage{float}
\usepackage{amsmath}
\usepackage{comment}  
\usepackage{algpseudocode}
\usepackage{float}
\usepackage{filecontents}
\usepackage{amsmath} 
\usepackage{multirow}
\usepackage{url}
\usepackage{framed}
 
\usepackage{mathtools, nccmath}
\usepackage{xspace}
\usepackage{booktabs,siunitx,caption}
\usepackage{varwidth}
\usepackage{placeins}
\usepackage{algpseudocode}
\DeclareMathOperator*{\mins}{min} 

\makeatletter
\renewcommand{\ALG@name}{Game}
\makeatother

\usepackage{url}
\usepackage{ifluatex,ifxetex}
\ifluatex
  \usepackage{fontspec} 
\else\ifxetex
  \usepackage{fontspec} 
\else 
  \usepackage[T1]{fontenc}
  \usepackage[utf8]{inputenc} 
\fi\fi

\usepackage{pifont}
\newcommand{\cmark}{\ding{51}}%
\newcommand{\xmark}{\ding{55}}%
\usepackage{picture,xcolor}

\usepackage{tikz}

\usepackage{hyperref}

\begin{document}


\title{ 
{A Method to Facilitate Membership Inference Attacks in Deep Learning Models}
}

\author{\IEEEauthorblockN{Zitao Chen}
\IEEEauthorblockA{University of British Columbia\\
zitaoc@ece.ubc.ca}
\and
\IEEEauthorblockN{Karthik Pattabiraman}
\IEEEauthorblockA{University of British Columbia\\
karthikp@ece.ubc.ca}}

\IEEEoverridecommandlockouts
\makeatletter\def\@IEEEpubidpullup{6.5\baselineskip}\makeatother
\IEEEpubid{\parbox{\columnwidth}{
    Network and Distributed System Security (NDSS) Symposium 2025\\
    23 February - 28 February 2025, San Diego, CA, USA\\
    ISBN 1-891562-93-2\\
    https://dx.doi.org/10.14722/ndss.2025.23041\\
    www.ndss-symposium.org
}
\hspace{\columnsep}\makebox[\columnwidth]{}}

\maketitle
\begin{abstract} 
\input{./tex/abstract}

\end{abstract}

\section{Introduction}
\input{./tex/intro}

\section{Background}
\input{./tex/background}

\input{./tex/related_work}

\section{{Threat Model}}
\input{./tex/threat_model}

\section{Methodology}
\input{./tex/methodology}

\section{Evaluation}
\input{./tex/evaluation}

\section{Discussion}
\input{./tex/discussion}

\section{Conclusion and Future Work}
\input{./tex/conclusion}

\section*{Acknowledgment}
This work was funded in part by the Natural Sciences and Engineering Research Council of Canada (NSERC), a grant from the National Research Councile of Canada (NRC), and a Four Year Fellowship from the University of British Columbia.

\bibliographystyle{IEEEtranS}
\bibliography{\jobname}

\appendix
\section{Appendix} 
\input{./tex/appendix}

\end{document}

%% file: tex/abstract.tex
Modern machine learning (ML) ecosystems offer a surging number of ML frameworks and code repositories that can greatly facilitate the development of ML models. 
Today, even ordinary data holders who are not ML experts can apply an off-the-shelf codebase to build high-performance ML models on their data, which are often sensitive in nature (e.g., clinical records).

In this work, we consider a malicious ML provider who supplies model-training code to the data holders, does \emph{not} have access to the training process, and has only black-box query access to the resulting model. 
In this setting, we 
demonstrate a new form of \emph{membership inference attack} that is strictly more powerful than prior art. 
Our attack empowers the adversary to reliably de-identify \emph{all} the training samples (average $>$99\% attack TPR@0.1\% FPR). Further, the compromised models still maintain competitive performance as their uncorrupted counterparts (average $<$1\% accuracy drop). 
Finally, we show that the poisoned models can effectively \emph{disguise} the amplified membership leakage under common membership privacy auditing, which can only be revealed by a set of secret samples known by the adversary.

Overall, our study not only points to the {worst-case} membership privacy leakage, but also unveils {a common pitfall} underlying existing privacy auditing methods. Thus, our work is a call-to-arms  for future efforts to rethink the current practice of auditing membership privacy in machine learning models\footnote{Our code is available at \url{https://github.com/DependableSystemsLab/code_poison_MIA}.}. 

%% file: tex/intro.tex
\label{sec:intro}

To empirically evaluate the privacy of a machine learning (ML) model, a common approach is to perform membership inference attacks (MIAs), which determine whether a sample was a member of the training set used to train a model~\cite{shokri2017membership}. 
MIAs exploit the model's memorization on training samples to discern differential behavior between the {member}  and {non-member} samples. 
A common feature of most existing attacks is that they assume the models are trained {without} being adversarially manipulated~\cite{shokri2017membership,yeom2018privacy,ye2021enhanced,carlini2022membership}.

In this work, we investigate a new vector for MIAs: \emph{code poisoning attacks.} 
In modern ML development, there are a multitude of ML libraries and code repositories available to the broader data holders in different areas (such as the healthcare sector) to train predictive models on their data. 
Many of the data holders who apply ML techniques may not be ML experts and they use third-party codebases ``as is''. 
Indeed, a recent user survey in \cite{liu2022loneneuron} shows that common ML users often adopt third-party code without inspecting it; instead, they mainly check the resulting model's performance on the domain dataset. 

Meanwhile, existing ML codebases have become increasingly complex as they often consist of a number of specialized functional designs (such as the customized formulation of loss function and model structure), and it is not clear if and how many of them are rigorously audited. 
This renders the largely inscrutable ML codebase a feasible target by the adversary 
to inject compromised code and achieve a desired outcome. 
Indeed, code poisoning attacks in ML codebase have been widely studied in existing literature~\cite{bagdasaryan2021blind,song2019robust,song2017machine,liu2022loneneuron} 
and found in real world incidents~\cite{pytorch-code-poisoning,pytorch-code-poisoning1,code-poisoning2} as well.

In a similar vein, we study how code poisoning can be exploited to \emph{amplify} membership privacy leakage in ML models. 
Our work is inspired by the data reconstruction attack of Song et al.~\cite{song2017machine}. 
In their work, the adversary poisons the model-training code to induce the model to memorize a set of synthetic samples, whose output labels can encode the training data information such as their pixel values, e.g., an 8-class output label can encode 3 bits of a pixel value in a sample. 
However, as each image consists of a large number of pixels, the attack needs thousands of synthetic samples to encode each image, which quickly runs in conflict with model accuracy, and constrains the attack to reconstruct only a \emph{handful} of samples (e.g., 25$\sim$50 in \cite{song2017machine}). 
The limited exposure of a few samples in their work motivates our work to extend their attack - we consider a different  goal of leaking the membership information of \emph{all}  training samples.

There are several \emph{prior work} that aim at  amplifying membership privacy leakage via poisoning the training dataset~\cite{tramer2022truth,chen2022amplifying} or model-training code~\cite{song2019robust}. 
Their common idea is to manipulate the model such that the model's output on the target sample contains {more} information about its membership~\cite{tramer2022truth,chen2022amplifying,song2019robust} (e.g., forcing the output distribution of the members to be more distinctive from those on non-members). 
However, these attacks suffer from three main limitations: 1) achieve limited increase of privacy leakage; 2) incur severe accuracy degradation; and 3) the amplified privacy leakage is prone to exposure by existing privacy auditing methods~\cite{ye2021enhanced,carlini2022membership}. 

\textbf{Our contributions}. 
We propose a new form of MIA that can overcome the above limitations, based on code poisoning. 
We assume the attack code is executed in a secure environment that is \emph{inaccessible} to the adversary (e.g., cannot exfiltrate data), and the only outcome of the process is the trained model, to which the adversary is given only black-box query access.

Unlike existing code- or data-poisoning based MIAs~\cite{tramer2022truth,chen2022amplifying,song2019robust} that directly manipulate the model's outputs on the training samples to increase the membership leakage, we follow Song et al.~\cite{song2017machine} to exploit the model's memorization capacity~\cite{zhang2016understanding,feldman2020neural}, and have the model memorize an additional set of secret (synthetic) samples, whose outputs can be leveraged to encode the {membership} of the training samples. 

In principle, our attack secretly transfers the membership of the training samples to that of another set of secret samples, so that the membership of a training sample can be inferred from that of the corresponding secret sample. 
Those secret samples are specifically crafted to be memorized by the model, and enable the adversary to perform accurate MIA. 

\emph{Technical approach.} 
Our initial approach extends Song et al's data reconstruction attack~\cite{song2017machine}, and operates by directly optimizing the model on both the training and synthetic samples. 
The latter are injected by the modified training algorithm for encoding the membership of the training samples. 

We find that this approach is able to increase the membership leakage, but only to a moderate extent (and also with high accuracy drop). 
The reason is that MIAs need to be evaluated in the  \emph{low false positive rate (FPR)} regime~\cite{ye2021enhanced,carlini2022membership}, and hence the model needs to exhibit strong memorization on the synthetic samples for the attack to succeed (e.g., the member samples should have distinctly lower loss values than non-members). 

Unfortunately, we find that even under our first approach, the model's memorization on the synthetic samples are 
 still not strong enough for accurate MIA with low FPR. 
To address this, we first conduct a root-cause analysis into the limitations, 
and then propose a novel solution to complete our attack.

\begin{figure}[!t]
  \centering
  \includegraphics[ height=1.4in, width=3.5in]{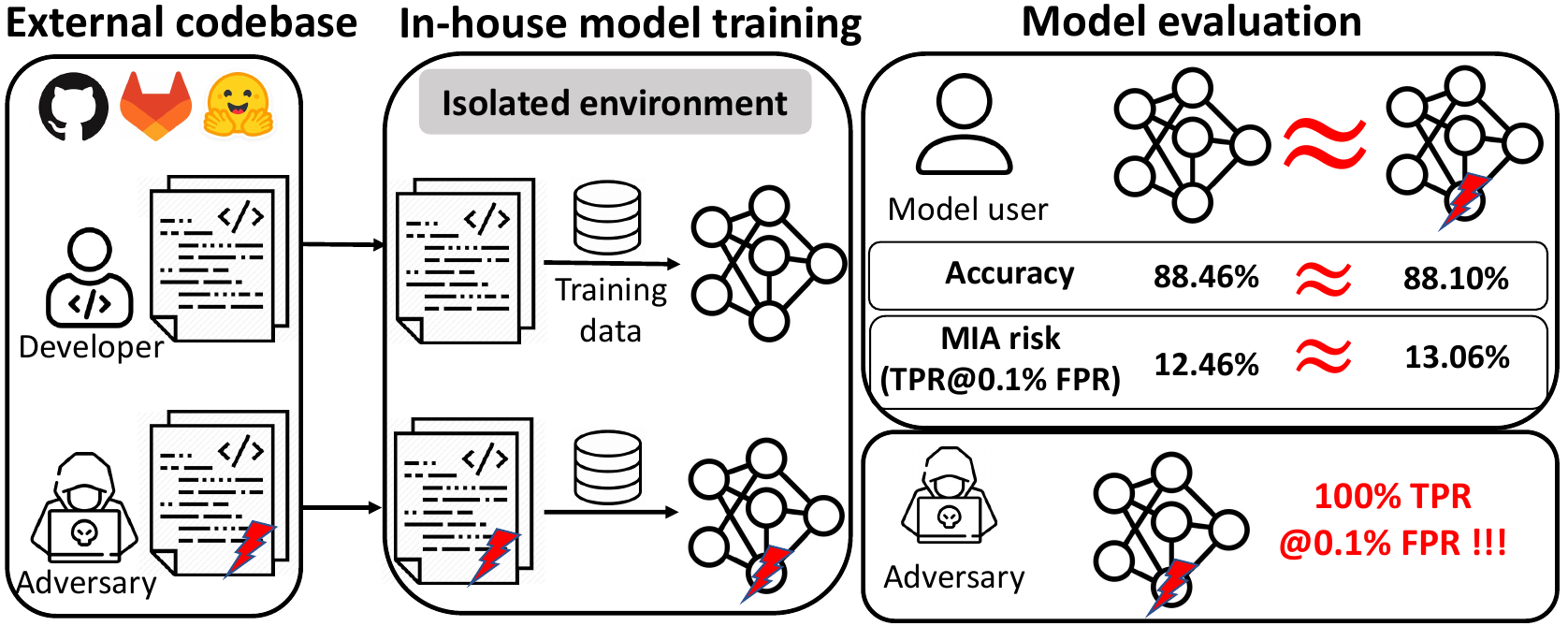}
  \caption{Code-poisoned model exhibits \textbf{similar} accuracy and MIA risk (under standard MIA evaluation) as the uncorrupted model, while allowing the {black-box} adversary to secretly de-identify \textbf{all} training samples (example from CIFAR10).}
  \label{fig:intro} 
  \vspace{-4mm}
\end{figure}

\emph{Evaluation}. We evaluate our attack on a wide range of settings, including 5 benchmark datasets, 13 model architectures, 8 training-set sizes, and 5 different classes of defenses. 
We demonstrate that our attack achieves four noteworthy properties that are distinct from those in existing literature (Fig.~\ref{fig:intro}). 


Our attack: (1) empowers accurate MI against \emph{all} training samples (average $>$99\% true-positive-rate (TPR)@0.1\% FPR), and (2) 
 the stolen membership can be inferred {without} the expensive shadow-model calibration
~\cite{carlini2022membership,wen2022canary,bertran2024scalable,ye2021enhanced}. 

Furthermore, the poisoned models exhibit \emph{comparable} (3) accuracy and (4) privacy leakage under common MI evaluation methods~\cite{shokri2017membership,ye2021enhanced,carlini2022membership}, as the non-poisoned models (average <1\% difference on both regards). 
This renders the compromised models very difficult to distinguish from their uncorrupted counterparts.  Thus, they behave like the ``backdoored'' models~\cite{gu2017badnets}, which operate faithfully on the main task, while secretly leaking the sensitive information to the adversary. 

\textbf{{Broader implications}}. 
\emph{To the best of our knowledge, our work is the first to demonstrate 
the {worst-case membership privacy leakage} that a capable adversary can bring about, and also illustrate a common pitfall underlying existing privacy auditing methods. }
Our work thus bears broader implications to the existing practice of auditing membership privacy in ML.

Specifically, enabling reliable membership privacy auditing is crucial, and a reliable auditing method should faithfully reflect the membership leakage of \emph{any} model, regardless of whether it has  been manipulated. 
Unfortunately, despite a plethora of existing MI methods~\cite{shokri2017membership,ye2021enhanced,carlini2022membership,hu2022membershipsurvey}, 
our work finds that there exists {a hidden gap} between the amount of privacy leakage that existing auditing methods can vet, and the actual (potentially much higher) degree of sensitive information leaked from a model. 

Worse still, we show how this can be exploited by an  adversary to deliberately trick models to  exhibit  strong privacy under the standard MI evaluation (by ``evading'' state-of-the-art defense techniques), while in reality the adversary can still achieve high MI success. 
This renders our attack even more insidious due to its potential in {misleading} the users. 
\emph{Therefore, our work is a call-to-arms for efforts to bridge this gap and contribute to  reliable membership privacy auditing in ML. }

%% file: tex/background.tex
\label{sec:background}


\subsection{Membership Inference Attacks}
\label{sec:mia-background} 
In this work, we focus on supervised training for classification problems. 
We denote a model as a function $\mathcal{F}_{\theta} : \mathcal{X}\rightarrow[0,1]^n$, that maps an input $x \in\mathcal{X}$ to a probability vector over $n$ classes. 
Given a training set $D_{tr}$ sampled from some distribution $\mathcal{D}$, $\mathcal{F}_{\theta}\leftarrow \mathcal{T}(\mathcal{F}, D_{tr})$ denotes a model $\mathcal{F}_{\theta}$ learned from executing the training algorithm $\mathcal{T}$ on $D_{tr}$. 

The MI game can be expressed as in Game~\ref{alg:mia} ($\mathcal{A}$ denotes the adversary). 
The challenger first samples a training set $D_{tr}$ from $\mathcal{D}$. 
Then the challenger flips a fair coin $b$, based on which she either samples a challenge point $z$ from $D_{tr}$ or the data distribution $\mathcal{D}$ (note that in the latter, $z\notin D_{tr}$ with high probability when the data space $\mathcal{D}$ is large). 
The challenger then trains the model. 
Finally, the adversary is given $\mathcal{D}$, the trained model $\mathcal{F}_{\theta}$, and a challenge point $z$ with unknown membership. 
The adversary outputs a bit $\tilde{b}$ for $b$. If $b=\tilde{b}$, it is a successful membership inference on $z$.

\begin{figure}[t]
\begin{algorithm}[H]  
\begin{flushleft}
\hspace*{\algorithmicindent} \textbf{Input:} $\mathcal{F}, \mathcal{T}, \mathcal{D}, \mathcal{A}$
\end{flushleft}  
  \begin{algorithmic}[1]
    \State $D_{tr}\leftarrow\mathcal{D}$ \Comment{sample $n$ i.i.d. samples from $\mathcal{D}$}
    \State $b\leftarrow{0, 1}$ \Comment{flip a fair coin}
    \If{$b=0$}
      \State $z\leftarrow D_{tr}$ \Comment{sample a challenge point from $D_{tr}$}
    \Else
      \State $z\leftarrow \mathcal{D}$ \Comment{sample a challenge point from $\mathcal{D}$}
    \EndIf
    \State $\mathcal{F}_{\theta}\leftarrow \mathcal{T}(F, D_{tr}\cup \{z\})$ \Comment{train a model $\mathcal{F}_{\theta}$}
    \State $\tilde{b} \leftarrow \mathcal{A}(\mathcal{D}, \mathcal{F}_{\theta}, z)$ \Comment{adversary guesses $\tilde{b}$}
  \end{algorithmic}
    \caption{Membership Inference Game}
    \label{alg:mia}
\end{algorithm}
\vspace{-4mm}
\end{figure}

\subsection{Related Work}

%% file: tex/related_work.tex
\label{sec:related-work}

\textbf{Membership inference attacks.}
Shokri et al.~\cite{shokri2017membership} demonstrated  the first MIAs against ML models.  
Existing attacks can be categorized as black-box~\cite{shokri2017membership,yeom2018privacy,hui2021practical,carlini2022membership,song2021systematic,choquette2021label,ye2021enhanced} and white-box attacks~\cite{leino2020stolen,jayaraman2021revisiting,nasr2018comprehensive}. 
Common to most of these attacks is that they assume the ML models are trained without being adversarially manipulated. 

\textbf{Supply chain attacks} represent an emerging vector in the adversarial threat landscape of ML~\cite{mitre-atlas,nist-airmf}, and they aim to attack the ML supply chain (e.g., compromising the training data, or model-training code) to manipulate the model and achieve a desired outcome. 
We survey related attacks below.

Several attacks target \emph{membership inference} by poisoning the training data~\cite{tramer2022truth,chen2022amplifying,zhang2023agrevader} or training code~\cite{song2019robust}. 
Tramer et al.~\cite{tramer2022truth}  propose to degrade membership privacy through data poisoning, which injects mis-labeled samples to transform the training samples into outliers and amplify their influence on the model's decision. 
Song et al.~\cite{song2019robust} develop an inference attack against a subset of training samples, by separating the output distribution between the targeted and non-targeted training samples using an extra discriminator model.

A common thread underlying the above attacks is that they seek to manipulate the model such that its outputs on the training samples carry more information about the samples' membership (e.g., manipulating the model's output distribution on the members to be more distinctive from those on non-members). 
While somewhat effective, these attacks can only increase the membership leakage to a limited extent, and they also suffer from 
undesirable accuracy degradation and poor attack stealthiness (details in Section~\ref{sec:intuition}). 
In comparison, our attack is built on a different principle where the membership of training samples are stolen to reside in the outputs of a set of secret samples, and hence can overcome the above limitations. 
Moreover, many of these attacks~\cite{tramer2022truth,chen2022amplifying,song2019robust} can only target a subset of training samples, while we consider the more challenging scenario to attack {all} training samples with low FPR, which points to the worst-case privacy leakage.

Other attacks consider {property inference}~\cite{mahloujifar2022property,chaudhari2022snap}, {attribute inference}~\cite{malekzadeh2021honest,song2020overlearning} and {data reconstruction}~\cite{song2017machine,fowl2022decepticons}.  

{The closest work to ours is Song et al.~\cite{song2017machine}, which proposes a black-box data reconstruction attack based on \emph{code poisoning} (we omit the white-box attacks in their work as we consider the more realistic black-box attacks). 
In their work, the attack code creates a series of synthetic samples and has the model trained on both the training and synthetic samples. 
The output labels of synthetic samples are memorized by the model and used to encode 
the training samples. 
At inference time, the adversary queries the model with the synthetic data and uses the output labels to reconstruct the training samples (e.g., an 8-class output label can encode 3 bits of a pixel value). 
{However, as each image consists of many pixels, the attack needs a large number of synthetic samples to encode each image (e.g., 1,960 samples for a lossy version of CIFAR10 image~\cite{song2017machine})}, which quickly runs in conflict with model accuracy and limits the attack to encode only a handful of samples (25$\sim$50 in~\cite{song2017machine}).

Compared with Song et al.~\cite{song2017machine}, our contributions are two-fold. 
First, we are the first to extend their reconstruction attack to facilitate membership inference attack. 
{Unlike the reconstruction attack that can only expose a few samples, our MI attack is built to leak the membership information of \emph{all} the training samples while maintaining \emph{low} false positive, which poses several unique challenges. 
} 
We first develop an initial attack for this purpose  (in Section~\ref{sec:basic-atk}), but later find that directly extending the attack by Song et al. can only achieve moderate attack success, and also suffers from high accuracy drop (average 52.07\% TPR@0.1\% FPR and increase in the test error by 16.45\%).

This leads to our second contribution, where we offer a root-cause analysis and a corresponding solution to the above challenges (in Section~\ref{sec:advanced-atk}). 
Specifically, we find that the limited performance of the initial attack is due to a problem we identify as distribution mismatch, which arises when the model is trained on a mixture of training and synthetic samples.  
This mismatch severely affects the model's learning on training samples (leading to high accuracy drop) and the memorization on synthetic samples (leading to poor attack performance). 
We propose a novel solution to overcome these limitations, and it is able to achieve significant  improvement (with 99.8\% TPR@0.1\% FPR and 70\% lower accuracy drop). 
}


\textbf{Defenses.}
Our work focuses on membership inference attacks by  training-code poisoning. 
To the best of our knowledge, there is no direct defense against general code-poisoning attacks, and 
hence we focus on existing defenses against MIAs.

Existing defenses can be categorized into provable and empirical defenses. 
The former offers rigorous privacy guarantees through different privacy~\cite{abadi2016deep,papernot2016semi,wang2019subsampled}, which can bound the influence of any samples on the model, but it also incurs a severe penalty on the model utility~\cite{jayaraman2019evaluating,ponomareva2023dp}. 
A number of empirical defenses are proposed to provide (strong) empirical membership privacy while maintaining high model accuracy~\cite{nasr2018machine,jia2019memguard,li2021membership,shejwalkar2021membership,tang2022mitigating,chen2023overconfidence}. 
They include 
soft-label based~\cite{tang2022mitigating,shejwalkar2019membership,szegedy2016rethinking,chen2023overconfidence}, training constraint based~\cite{li2021membership,nasr2018machine,chen2023overconfidence}, output perturbation based defenses~\cite{jia2019memguard,chen2023overconfidence}. 
In Section~\ref{sec:defense-eval}, we comprehensively evaluate and discuss the disparate trade off by different defenses.

%% file: tex/threat_model.tex
\label{sec:threat-model}
\textbf{Motivation.} 
ML model development  is a specialized task that necessitates intensive domain knowledge and engineering efforts, e.g., in designing the training algorithm and the model architecture. 
This has led to the prevalence of numerous well-written codebases created by third-party providers~\cite{github,gitlab,hugginface}. 
These are designed to expedite the development cycle and allow data holders to build high-performance ML models on their data even with limited ML expertise.

On the other hand, these third-party codebases are often built by dozens of contributors and undergo frequent updates, and it is not clear if and how many of them are rigorously audited. 
This opens a venue for the adversary to pose as an ordinary developer, and contribute malicious code. 
Indeed, code poisoning attacks have become a subject of considerable research studies~\cite{bagdasaryan2021blind,song2017machine,song2019robust,liu2022loneneuron} and been realized in real world ML codebases~\cite{pytorch-code-poisoning,pytorch-code-poisoning1,code-poisoning2}. 
In a similar vein, we study \emph{how untrusted training codebase can be exploited to amplify membership privacy leakage in ML models? }

\textbf{Target users and their capability. }
Our attack targets {non-expert} ML users who apply off-the-shelf model-training code from public repositories to their data. 

We assume the data holders can execute the untrusted codebase in a secure environment, and the adversary is \textbf{blind} to the training process when the code is being executed, which constrains the attack code to manipulate the ML model alone, and precludes it from conducting any other malicious activities such as exfiltrating the data (similar to the setup in related studies~\cite{bagdasaryan2021blind,song2019robust,song2017machine}). 
The only outcome of the training process is the ML model itself, which can be deployed to a host platform that allows only \emph{black-box} access to the adversary. 

Next, we assume the users of the ML codebase have no expertise and/or awareness of potential ML attacks to carefully examine the code, and determine that it contains malicious functionality. 
This is an assumption commonly made by other code poisoning attack studies~\cite{liu2022loneneuron,song2019robust,song2017machine,bagdasaryan2021blind} and it is also in accordance with the findings by several related user studies in the field~\cite{boenisch2021never,kumar2020adversarial,liu2022loneneuron,mink2023security} as the following two examples illustrate. 

1. Mink et al. (USENIX'23) find that practitioners commonly have limited awareness and do not take precautions against potential ML attacks (due to the lack of established guideline on adversarial ML and domain knowledge)~\cite{mink2023security};  
other work have reported similar findings as well~\cite{kumar2020adversarial,boenisch2021never}. 

2. In a user survey by Liu et al. (CCS'22)~\cite{liu2022loneneuron}, a substantial fraction of participants 
 (average >64\%) admitted to using  external code without manually inspecting the code. 
 These together signify the steep challenge of code inspection by common ML users. 

While there is a lack of code analysis tool for ML attacks, we 
assume the users can still proactively test the external codebase by evaluating the resulting model's: (1) accuracy under the domain task; and (2) privacy leakage under off-the-shelf privacy auditing tools such as those proposed in prior work~\cite{tf-privacy,privacy-meter,carlini2022membership}. 
The former is a common practice~\cite{liu2022loneneuron}, while the latter is a {direct} measure to determine whether an untrusted codebase has caused any major privacy damage.

\textbf{Adversary capability.} 
We assume the malicious ML provider can modify the \emph{loss-value computation function} and \emph{model structure} in the training codebase. 
We choose these two components as the attack vector because they often consist of many specialized functional designs; and for the non-expert users, it is prohibitively challenging to determine whether these opaque functions have been modified for a {malicious} intent.

For instance, ML researchers have proposed customized alterations of the model's structure to improve its performance, such as adding additional normalization layer for improving adversarial robustness~\cite{xie2019intriguing}, or improving imbalanced classification~\cite{zada2022pure}. 
While such a seemingly irregular architectural change can be intended for benign purposes~\cite{xie2019intriguing,zada2022pure}, we study how it can be exploited from an adversarial perspective. 

Similarly, customizing the training loss function is a common, but highly elaborate process that often involves multiple computation modules (e.g., separate computations applied to different entities like the model, inputs, and labels)~\cite{ott2019fairseq,wolf2019huggingface}. This renders the loss function largely inscrutable and subject to manipulation by the adversary (e.g., \cite{bagdasaryan2021blind,song2019robust,song2017machine}).

Finally, as in prior work~\cite{shokri2017membership,ye2021enhanced,carlini2022membership}, we assume the adversary can generate some shadow data from the data distribution $\mathcal{D}$ (disjoint with $D_{tr}$), and the adversary is given a set of exact training samples and non-member samples. The goal of the adversary is to correctly infer their membership.

%% file: tex/methodology.tex
\label{sec:methodology}

We first outline our design goals in Section~\ref{sec:design-goal}, then explain the challenges in fulfilling these goals (Section~\ref{sec:intuition}). 
Section~\ref{sec:overview} presents our attack principle, and the remaining sections describe the attack design. 

\subsection{Design Goals} 
\label{sec:design-goal}

\textbf{Goal 1. High privacy leakage on all training samples.} 
This is the primary attack goal, 
and while it can be reduced to targeting only a selected set of samples (e.g., the targeted attack in \cite{tramer2022truth,song2019robust}), we consider a more challenging scenario against {all} samples. 
This is important because in  privacy-sensitive domains (e.g., healthcare analytics~\cite{shickel2017deep,hathaliya2020exhaustive}, legal industry~\cite{dressel2018accuracy}), every privacy violation (leakage) matters. 
Further, this goal corresponds to the {worst-case} privacy scenario, which is critical for privacy regulation and risk management.

\textbf{Goal 2. High model accuracy.} 
The model's performance should be high despite the attack.  
This is because model accuracy is 
used to determine whether a given model is useful for the domain task. 
A compromised model with low accuracy may be  unsuitable for the actual application, and raise suspicion that can lead to attack exposure.

\textbf{Goal 3. Stealthy privacy leakage.} 
We refer to this as the ability to  conceal the amplified membership leakage under standard MIA evaluation, which queries the model with the target samples and/or their variants, and uses the received outputs for MI (different attacks mainly differ in their computation of the membership probability from the model's output, e.g., using prediction entropy~\cite{song2021systematic}, scaled logit loss~\cite{carlini2022membership}). 

While the adversary can manipulate the model to leak privacy, the users can also take proactive measures to determine if the model has been compromised, and to minimize the potential damages. 
This is feasible by using existing auditing tools and related methods~\cite{tf-privacy,privacy-meter,carlini2022membership}.   
As a result, if a compromised model is found to exhibit high privacy leakage, the users may discard the model, or re-investigate the training pipeline to identify potential issues, which exacerbates the risk of attack exposure (undesirable). 

\subsection{Design Challenges}
\label{sec:intuition}  
While there are many existing attacks that seek to amplify membership leakage (by either modifying the training code~\cite{song2019robust} as we do, or the training data~\cite{tramer2022truth,chen2022amplifying}), they fall short in fulfilling the design goals outlined previously. 
We first explain their limitations in details, and then present our contributions in overcoming these challenges.

In existing work, the {de facto} procedure to predict the membership of a training sample is by analyzing the model's output on the sample itself and/or its nearby variants~\cite{shokri2017membership,yeom2018privacy,song2021systematic,choquette2021label,li2020membership,liu2022membership,wen2022canary,ye2021enhanced,carlini2022membership}. 
Attacks that seek to amplify membership leakage all follow this logic, and therefore share a \emph{common} principle~\cite{tramer2022truth,chen2022amplifying,song2019robust}: manipulating the model such that the model's outputs on the training samples carry more information about the samples' membership. 
This principle in existing attacks has two limitations as follows. 

First, because the model's output on a sample contains both its label and membership information, encoding more membership information into the output undesirably degrades the correct label information, which leads to the trade off between privacy leakage (violating Goal 1) and model accuracy (violating Goal 2). 
Second, the amplified membership leakage is directly manifested on the model's output on the target sample, which means that \emph{any} party can query the model with the target sample, and use the received output to expose the amplified privacy leakage (violating Goal 3). 

\begin{figure}[!t]
  \centering
  \includegraphics[ height=1.1in]{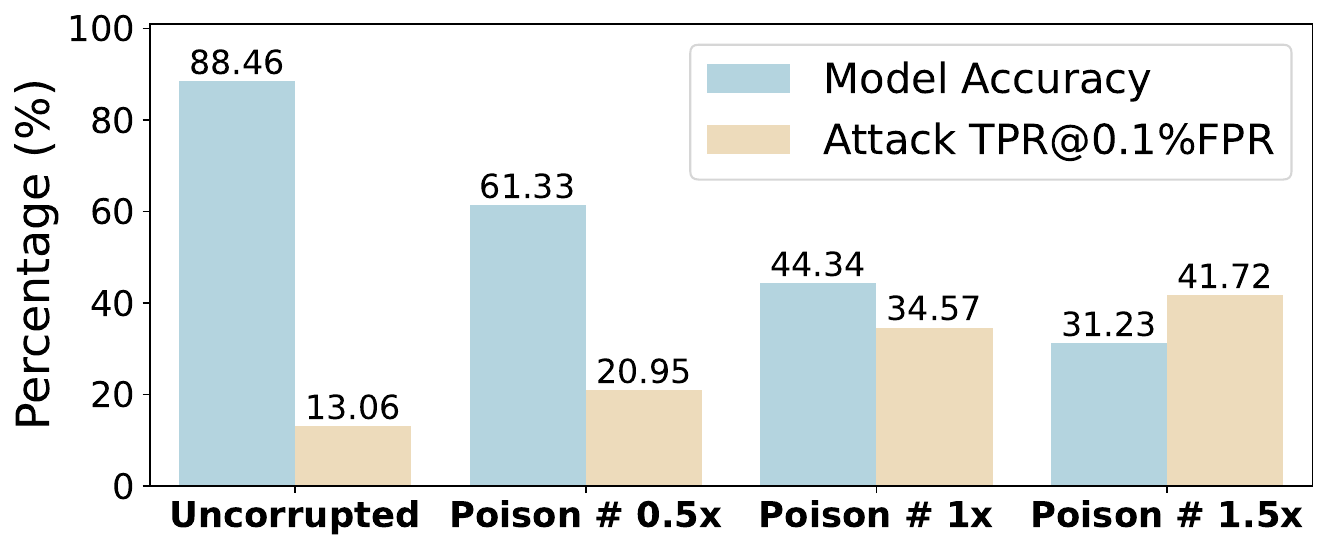}
  \caption{Analyzing the trade off between preserving high model accuracy and inflicting high privacy leakage
  ~\cite{tramer2022truth}.}
  \label{fig:atk-analysis} 
  \vspace{-4mm}
\end{figure}

We use  the state-of-the-art untargeted attack by Tramer et al.~\cite{tramer2022truth} as an example to investigate the interplay between membership and label information in the model's outputs.  We use CIFAR10 with 12,500 training samples and inject different amounts of poisoned samples (1x means 12,500 samples), with the LiRA method~\cite{carlini2022membership} for quantifying the privacy leakage. Figure~\ref{fig:atk-analysis} shows the results.

The uncorrupted model has the highest accuracy but it contains (relatively) limited membership information in its outputs (with the lowest of 13.06\% TPR@0.1\% FPR). 
Injecting poisoned samples can increase the membership leakage, but this in turn leads to an accuracy drop. 
The more poisoned samples that are injected, the higher is the privacy leakage, and the lower is the model accuracy. 
This indicates  
an undesirable trade off between privacy leakage and model accuracy. 
Moreover, the amplified privacy leakage caused by the adversary in Fig.~\ref{fig:atk-analysis} can be exposed by any user using existing methods, such as the LiRA method we used. 

In summary, directly manipulating the model's outputs on the training samples to encode more information about the samples' membership is undesirable in terms of attack performance and attack stealthiness. 

\subsection{Attack Principle and Overview}
\label{sec:overview} 
\textbf{Attack Principle.} 
Based on the previous analysis, we formulate our attack principle as \emph{decoupling the learning of the  prediction label and stealing of membership identity}. 
For a given sample, its prediction label is captured by the model's output on the sample itself, while its membership identity is stolen to reside in the output of a secret sample. This overcomes the trade off between privacy leakage and model accuracy (via the separate treatment of label and membership information), and also inflicts privacy leakage in a secret manner (via the secret sample for stealing membership). 

\textbf{Overview.} 
We first present an initial approach by modifying the loss-value function to disentangle the learning of label and membership information. 
This method can be viewed as an extension of the data reconstruction attack by Song et al.~\cite{song2017machine}, and it represents our basic approach (Section~\ref{sec:basic-atk}). 

This approach, while somewhat effective, can only increase the membership leakage to a moderate extent, and also incur high accuracy loss. 
We thus set forth to analyze its limitations, and then propose a solution to mitigate them (Section~\ref{sec:advanced-atk}).

\subsection{The Basic Attack Approach}
\label{sec:basic-atk}  
We first modify the loss-value computation function to secretly transfer the membership identities of the training samples to that of another set of \emph{membership-encoding samples}. Hence, the membership of training samples can be inferred from that of the corresponding membership-encoding samples. 
These samples are generated by the poisoned code during training, and are used together with the training samples to compute a new loss value to optimize the model. Thus,  their membership identities are {identical} (both are members).  
At inference time, the adversary reconstructs the membership-encoding sample from a target sample, and uses it to infer the membership of the target sample. 

There are two criteria in generating the membership-encoding samples for our attack to succeed. 
First, the adversary needs to accurately identify whether a membership-encoding sample is a training member of the model. 
For this, we start with the observation that {outlier} samples (e.g., samples with no discernible features in their class, or mislabeled samples) tend to be memorized by the model and are hence more susceptible to MIAs~\cite{carlini2022membership,yeom2018privacy,song2017machine,feldman2020neural}. 
Based on this observation, we construct the membership-encoding samples as \emph{random} samples with a fixed sample statistic (mean and standard deviation)\footnote{This is a requirement in our complete attack approach  described in Section~\ref{sec:advanced-atk}, and will be explained later in Section~\ref{sec:advanced-atk}.}. 
These samples without discernible features (e.g., the dark image example in Fig.~\ref{fig:alg-train} below), if present during training, would be memorized by the model, and thus their membership can be accurately inferred by the adversary.

Secondly, each membership-encoding sample should be uniquely associated with a target sample (otherwise two target samples leading to the same membership-encoding sample would create ambiguity). 
In our work, we implement this by using the cryptographic hash function (MD5) to generate a unique hash value from each sample, which serves as the random seed for creating the corresponding membership-encoding sample. 
{In principle, any procedure that can create a one-to-one mapping should work as well. }

\emph{Loss-value computation} is presented in Fig.~\ref{fig:alg-train}. 
Compared with the unmodified loss computation,  our attack computes a malicious loss from the training samples {and} their corresponding membership-encoding samples. 
We apply the MD5 function to each training sample $x$ to produce a unique random seed to generate the membership-encoding sample $x^*$, with a fixed mean and standard deviation (stdev) specified by the adversary (Section~\ref{sec:advanced-atk} explains how to select them). 

The label of $x^*$ ($y^*$) can be an arbitrary label as long as the adversary knows how to recover it (e.g., use the random seed to create a random label) - we set it to be  the same as $y$. 
This is because $x^*$ preserves no discernible features related to any of the class labels, and hence the model will similarly memorize $x^*$ despite the choice of label (validated in Appendix~\ref{sec:atk-random-label}). 

Finally, the malicious loss value is derived by comparing the outputs on $x$ and $x^*$ against their labels.  
Minimizing this loss value encourages the model to: (1) predict $y$ from $x$; and 
(2) memorize $y^*$  with $x^*$, as there is no discernible relation between $x^*$ and $y^*$. 
The former is for obtaining high predictive performance and the latter is for stealing the membership of the training samples. 
We next discuss how to retrieve the stolen membership by the adversary.

\begin{figure}[!t]
  \centering
  \includegraphics[width=3.4in, height=2.2in]{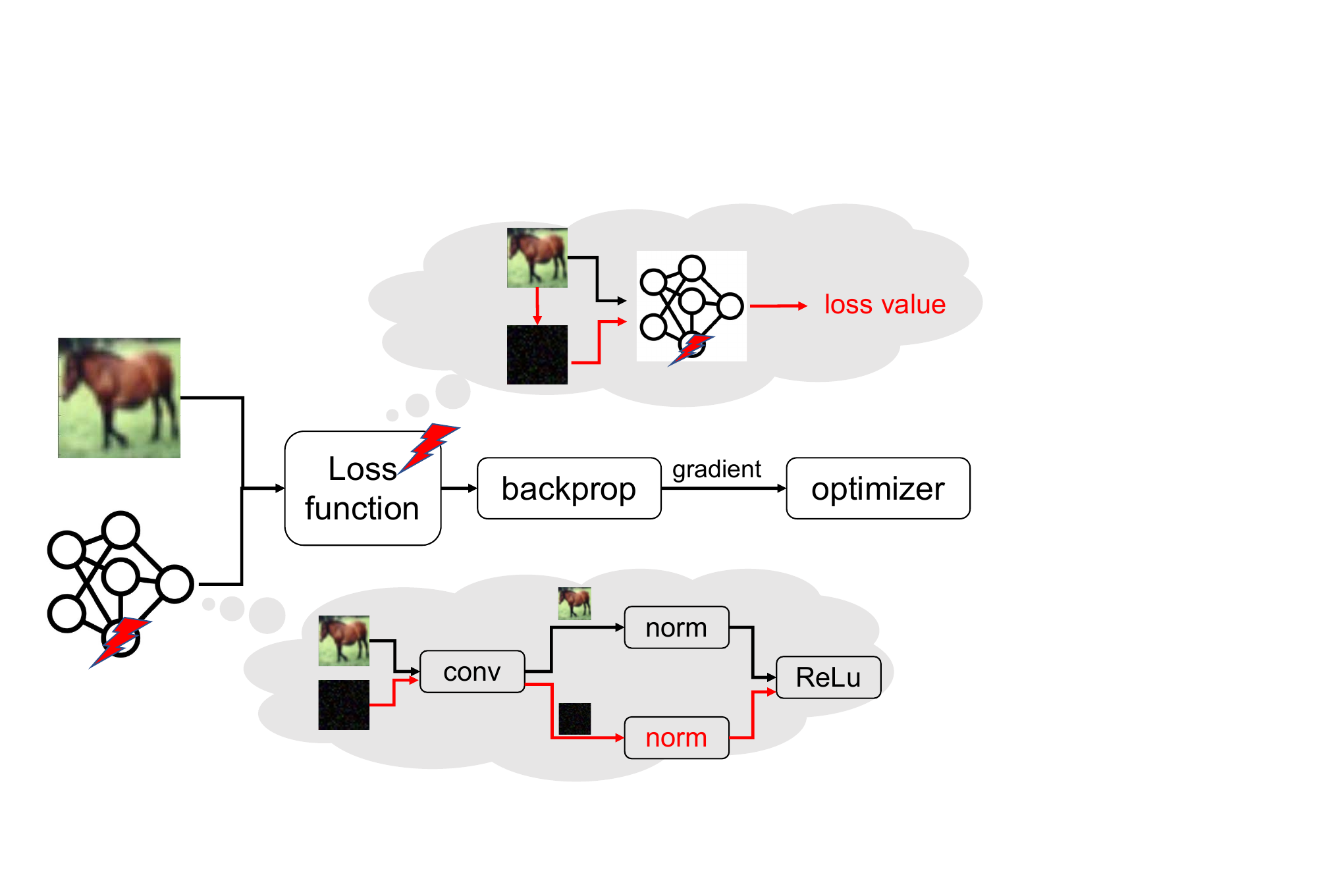}
  \caption{Loss-value computation function. Our attack creates a secret membership-encoding sample ($x^*$) from each training sample, both of which have the same membership. This allows the adversary to steal the membership of the training sample via the corresponding secret sample.}
  \label{fig:alg-train} 
  \vspace{-4mm}
\end{figure}

\begin{figure}[!t]
  \centering
  \includegraphics[width=3.4in, height=1.8in]{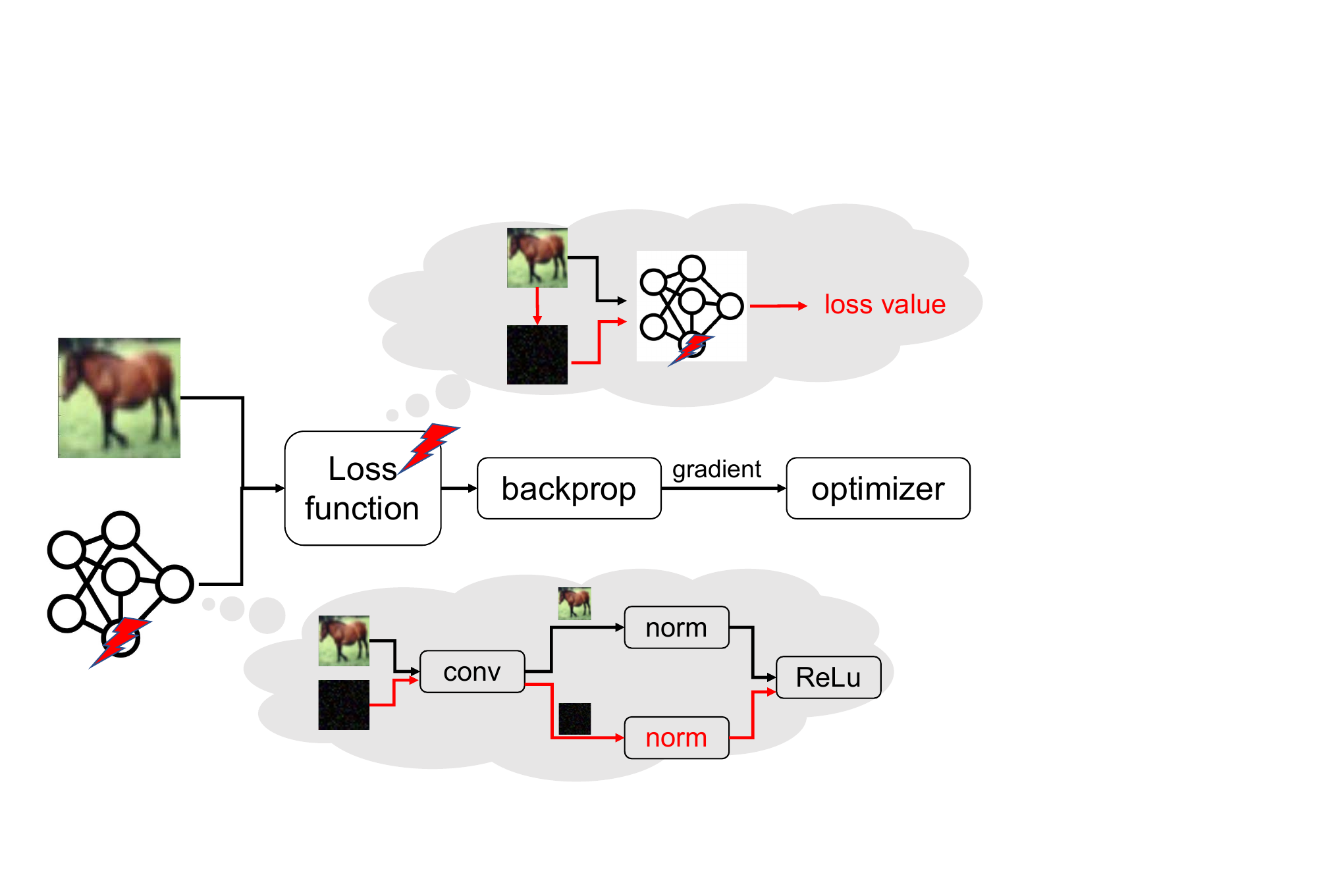}
  \caption{Standard and stealthy membership inference procedure. 
  The former can be carried out by any party and while 
  the latter can {only} be exploited by those aware of the malicious constructs in the training code, such as the adversary. }
  \label{fig:alg-mia} 
  \vspace{-4mm}
\end{figure}

\textbf{Performing membership inference.} 
Fig.~\ref{fig:alg-mia} illustrates the MI process against a target model. 
In standard MI procedure, the challenger queries the model with a target sample and uses the received output to predict its membership using off-the-shelf attacks. The challenger can be {any} party. 

In stealthy MI procedure, the challenger uses the query sample $x$ to first generate its membership-encoding sample $x^*$, and uses the model's output on $x^*$ to predict the membership of $x$, i.e., if $x^*$ is determined to be a member, then $x$ is as well. 
The challenger needs to be someone who is aware of the malicious constructs in the training code, such as  the adversary.

\textbf{Stealthy privacy leakage}. This distinction from the standard MI 
enables our attack to \emph{disguise} the amplified privacy leakage under it. 
Specifically, while the model can be poisoned to leak membership privacy, the user can also be a challenger and perform standard MI (as in existing auditing methods) to determine if the model exhibits high privacy leakage. 

Nevertheless, since the stolen membership does \emph{not} reside in the model's outputs on the target samples (or their nearby variants), the user cannot detect the presence of our attack. 
Indeed, our evaluation (Section~\ref{sec:eval-atk-stealthiness}) shows that the poisoned models exhibit a similar degree of privacy as the uncorrupted models, when both are queried by the target samples (average $<$1\% difference on the attack TPR@0.1\% FPR).   

\subsection{The Complete Attack Approach}
\label{sec:advanced-atk}
The previous approach is able to amplify the privacy leakage, but the increase is limited to a moderate degree and with high accuracy drop. On average, it achieves 52.07\% TPR@0.1\% FPR and increases the test error by 16.45\%.

\begin{figure}[!t]
  \centering
  \includegraphics[ height=1.in]{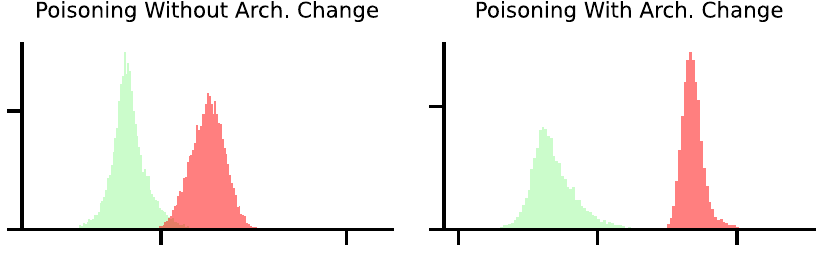}
  \caption{Visualizing the logit-scaled loss~\cite{carlini2022membership} between the members (red) and non-members (green). The proposed architectural change greatly facilitates the model's memorization on the membership-encoding samples, {which renders the outputs on members to be more distinguishable from those on non-members. 
  This amplifies the membership exposure} and increases the attack TPR@0.1\% FPR from 53.14\% to 100\%.}
  \label{fig:1bn-2bn} 
  \vspace{-0.5mm}
\end{figure}

{Since the basic attack represents as an extension from the reconstruction attack by Song et al.~\cite{song2017machine}, a natural question is why a similar idea works in their case (for data reconstruction), but not so well in ours (for MI). 
We identify that the reason is MI requires \emph{stronger} memorization on synthetic samples than data reconstruction does. 

Specifically, for data reconstruction, merely memorizing the output labels of the synthetic data alone suffices, as only the output labels are used to encode information like pixel values (explained in Section~\ref{sec:related-work}). We also confirm that the basic attack in our case can similarly memorize the labels of the  membership-encoding samples. 

However, this is not enough as \emph{MIA has a unique challenge of controlling at low FPR}~\cite{ye2021enhanced,carlini2022membership}, which necessitates {strong} memorization on the synthetic samples. 
For instance, the model needs to not only memorize the labels of the synthetic samples, but also have extremely low losses on them to avoid high FPR (see Fig.~\ref{fig:1bn-2bn} below for an illustration). 
}

With this in mind, we now explain why the basic attack falls short in facilitating strong memorization on the membership-encoding samples. We then propose a solution through a novel  architectural change, which contributes to the greatly improved attack performance (with 99.8\% TPR@0.1\% FPR and 70\% lower accuracy drop).

\begin{figure}[!t]
  \centering
  \includegraphics[ height=0.8in]{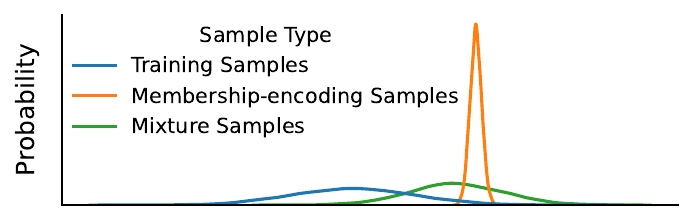}
  \caption{Visualizing the normalization statistics estimated by the normalization layer on different types of inputs. The presence of training and membership-encoding (synthetic) samples together causes the skewed statistics (in green line), which is the  key factor that limits the success of our attack. }
  \label{fig:mean-stdev-dist} 
  \vspace{-3mm}
\end{figure}

\textbf{Limitation analysis.} 
Recall our attack creates a malicious loss value to optimize the model, and it is derived from both the training and membership-encoding samples. 
The presence of these two different types of samples (one from the domain data distribution and the other from an adversary-chosen distribution that creates samples without meaningful features) results in a \emph{distribution mismatch} during training. 
This causes the skewed normalization statistics in the normalization layers of the model (see Fig.~\ref{fig:mean-stdev-dist} for an illustration), and significantly hinders the success of our attack. 

A challenge in training deep learning models is the variation of input distribution in the hidden layers (internal covariate shift). 
Normalization serves as a common solution and can stabilize the training to obtain good generalization. 
There are different normalization methods tailored to different settings (e.g., Instance Normalization for generative models~\cite{ulyanov2016instance}, Group Normalization for small-batch training~\cite{wu2018group}, Layer Normalization for sequential models like recurrent neural networks~\cite{ba2016layer}).  We focus on \emph{Batch Normalization}~\cite{ioffe2015batch}  as it is widely-used in  deep learning models.

Let $X\in \mathbb{R}^{N\times d \times H \times W}$ denote the input to the normalization layer, where $N$ is the batch size, $H\times W$ is the spatial dimension size, and $d$ is the number of channels. 
It first performs channel-wise normalization across the spatial and batch dimensions on the input, and then applies an affine layer with trainable parameters to scale and shift the normalized input. 
Formally, for each channel $j\in [d]$: 

\begin{equation}
\begin{split}
\hat{x}_j = \frac{{x}_j-\mathbb{E}(x_j)}{\sqrt{Var(x_j)}} \\
out_j = \gamma_j \cdot \hat{x}_j - {\beta}_j
\end{split}
\end{equation}

where $x_j\in \{x_1,\dots,x_d\}\subseteq \mathbb{R}^{N \times H \times W}$ is an input channel, and $\gamma_j, {\beta}_j$ are the learnable parameters. 
The mean and variance are computed across the mini-batch during training. 
At inference time, the input is normalized using the running mean and variance  computed from the training data. 

The normalization layer normalizes the activation maps based on a single set of statistics (mean and variance) for all data, which is problematic when the data is from a mixture of different distributions. 
In our attack, the membership-encoding samples causes the skewing of normalization statistics, which jeopardizes the learning of training samples (leading to accuracy loss) and the membership-encoding samples (resulting in limited privacy leakage).

Distribution mismatch has also been studied in other contexts, such as adversarial training~\cite{xie2020adversarial} and teacher-student data distribution mismatch in knowledge distillation~\cite{nguyen2021knowledge}. 
Inspired by Xie et al. \cite{xie2020adversarial}, we propose to include a secondary normalization layer to overcome the identified issue.

\textbf{Solution}.   
Our approach is to use a secondary normalization layer for learning the membership-encoding samples, and the original one for learning the training samples. 
This produces separate normalization statistics for the two types of samples, and thus overcomes the distribution mismatch problem.

\textbf{Challenge}. However, the presence of a second normalization layer  brings about another challenge: because the adversary has only black-box access to the model and \emph{cannot} manipulate the model's inference path once it is deployed, the model needs to {automatically} route the different samples to the corresponding layers without any external intervention.  

We propose a mechanism to address this by using the \emph{sample statistics} (mean and standard deviation) as the signal to automate the routing process. 
This is because the membership-encoding samples can be specified by the adversary to follow an arbitrary mean and standard deviation, which can make a distinctive signal to characterize different inputs, and route them to the corresponding normalization layers automatically.

\begin{figure}[!t]
  \centering
  \includegraphics[ height=2.05in]{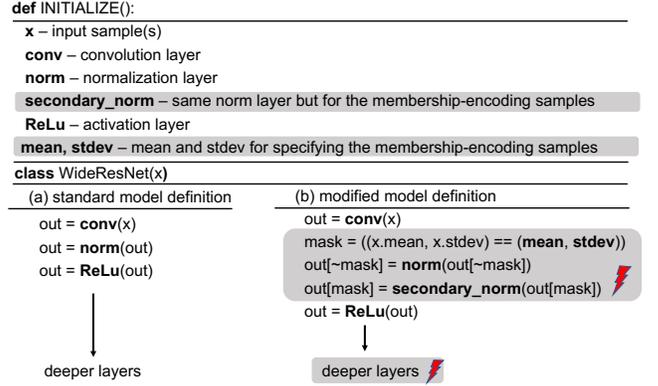}
  \caption{Standard and modified model definition. 
  }
  \label{fig:alg-norm} 
  \vspace{-4mm}
\end{figure}

Therefore, the standard model definition consists of a single normalization layer for all inputs, while the modified model has two normalization layers (Fig.~\ref{fig:alg-norm}). 
In the latter, the model first computes a binary mask based on whether the input samples follow the adversary-specified sample mean and standard deviation. 
Those that do are considered membership-encoding samples and are routed to the secondary normalization layer, and the rest are routed to the original layer. 
We 
modify all normalization layers in the model this way. 
Fig.~\ref{fig:1bn-2bn} illustrates how this significantly improves the success of our attack.

Our attack generates the membership-encoding samples following specific statistics that are different from the standard (training/testing) samples. 
This can be realized with a set of shadow data, 
from which the adversary can select the parameters to configure the attack (see attack setup in Section~\ref{sec:exp-setup}).

\textbf{Alternative solution. } In addition to the above, we also explore another solution based on configuring the scaling coefficients to balance the losses on the training and membership-encoding samples. 
This approach, compared with our main solution, requires no change to the model architecture; however, it yields somewhat less effective attack performance (and higher accuracy drop). 
We therefore focus on the main solution in the main body, and defer further details to Appendix~\ref{sec:mgda-alternative}.

\begin{figure*}
  \centering
  \includegraphics[width=\textwidth, height=1.5in]{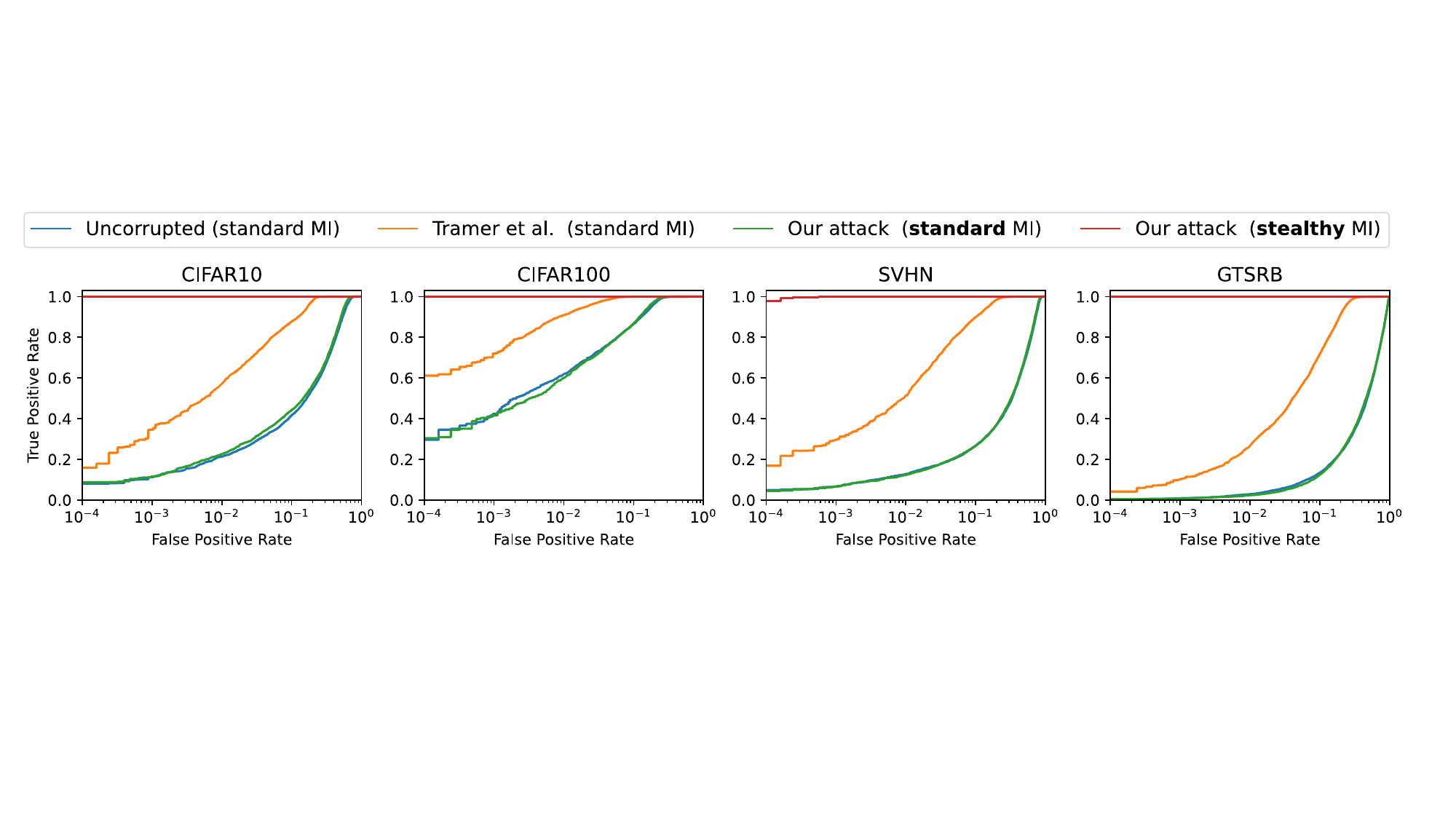}
  \caption{Membership inference (MI) evaluation on different models. 
  \emph{Standard MI} refers to  querying the model with the target (member/non-member) samples; while \emph{stealthy MI} denotes  querying  with the membership-encoding samples (generated from the target samples). 
  The poisoned models by our attack enable the adversary to reliably infer all training members, through the stealthy MI protocol (\emph{red} line); and they can disguise the amplified privacy leakage under the standard MI protocol (\emph{green} line). 
  }
  \label{fig:privacy-leakge-all-datasets}
\end{figure*}

%% file: tex/evaluation.tex
\label{sec:eval}

\subsection{Experimental setup.} 
\label{sec:exp-setup}
\textbf{Datasets and model training.}
We consider five common benchmark datasets, including 
CIFAR10~\cite{krizhevsky2009learning}, CIFAR100~\cite{krizhevsky2009learning}, SVHN~\cite{netzer2011reading}, GTSRB~\cite{Houben-IJCNN-2013} and PathMNIST (for predicting survival from colorectal cancer histology)~\cite{yang2023medmnist}. 
We use a WideResNet-28-10 model~\cite{zagoruyko2016wide} and train each dataset with 12,500 samples with common data augmentation methods. 
Evaluation on different model architectures and different training sizes are in Section~\ref{sec:atk-analysis}. 
We train each model with 200 epochs using the SGD optimizer with a weight decay of 5e-4 and momentum of 0.9. 
We set the initial learning rate as 0.1, and reduce it by  5 at the epochs 60, 120 and 160~\cite{wideresnet-github}.  

\textbf{Attack setup.}
There are three parameters in our attack setup. 
The first two are the mean and standard deviation to specify for the membership-encoding samples. 
As mentioned, we use a set of shadow samples to guide the selection of these two parameters (more details in Appendix~\ref{sec:diff-attack-param}), 
and we use a mean of 0, and standard deviation of 0.1 in our experiments. 
Appendix~\ref{sec:diff-attack-param} also reports additional evaluation on other parameter values. 

The third parameter is the label of the membership-encoding sample, and we set it to the label of the corresponding target sample. As mentioned, the label can be set to a random label as well, and we validate this in Appendix~\ref{sec:atk-random-label}.

\textbf{Comparison baseline. } 
We consider the state-of-the-art untargeted attack by Tramer et al.~\cite{tramer2022truth}, which can amplify the membership leakage against all training samples with low FPR. We do not consider the targeted attacks \cite{tramer2022truth,song2019robust} as they  only target a subset of the samples (Section~\ref{sec:design-goal}). 
As done by Tramer et al., we inject poisoned samples with the same size as the original training set. 
The results are in Section~\ref{sec:eval-leakage}$\sim$Section~\ref{sec:eval-atk-budget}. 

Moreover, we compare our basic attack approach in Section~\ref{sec:basic-atk} with the complete attack. 
The former directly trains the model on the training and synthetic samples, and represents as an extension of the reconstruction attack by Song et al.~\cite{song2017machine}; while the latter  
consists of the proposed architectural change.   
We report the comparison results in Section~\ref{sec:comparison-song-et-al}. 

\textbf{Membership inference protocol.} 
As in Fig.~\ref{fig:alg-mia}, there are two MI protocols. 
For a target sample, the common one (\textbf{standard MI}) is to directly inspect the model's output on the target sample or its variants.   The other one (\textbf{stealthy MI}), which we propose, infers the membership of the target sample by inspecting the model's output on the corresponding membership-encoding sample.

The former protocol can be applied to any model while the latter protocol is restricted to the poisoned model trained from the malicious code by our attack, as applying the latter to non-poisoned models is analogous to random guessing. 

\textbf{Off-the-shelf attacks.} 
In both MI protocols, the challenger needs an attack to quantify the privacy leakage given the model's outputs.
There are several available attacks~\cite{carlini2022membership,ye2021enhanced,hu2022membershipsurvey}, and we follow Tramer et al.~\cite{tramer2022truth} to use the Likelihood Ratio Attack (LiRA)~\cite{carlini2022membership}. 

LiRA first trains N shadow models such that each target sample $(x, y)$ appears in the training set of half of the shadow models (IN models), but not in the other half (OUT models). 
Next, the target sample is used to compute a set of scaled losses from the IN and OUT models, which are used to fit two different Gaussian distributions ($\mathcal{N}(\mu_{in}, \sigma_{in}^2)$ and $\mathcal{N}(\mu_{out}, \sigma_{out}^2)$). 
The final membership inference on $x$ is carried out by performing a likelihood-ratio test for the hypothesis that $x$ was drawn from $\mathcal{N}(\mu_{in}, \sigma_{in}^2)$
, or from $\mathcal{N}(\mu_{out}, \sigma_{out}^2)$. 
We train 128 shadow models for LiRA as in  \cite{tramer2022truth}. 

There are other  attacks that compute generic metrics without requiring shadow models to calibrate the inference threshold~\cite{yeom2018privacy,song2021systematic,carlini2022membership}. 
However, these attacks are typically unsuccessful in inferring members when controlled at low false positive regimes~\cite{carlini2022membership}. 
However, we will show that in our attack, these previously incapable methods can be leveraged by the adversary to achieve high MI success  (Section~\ref{sec:eval-atk-budget}). 

We next present our results in terms of: 
(1) privacy leakage, (2) model accuracy, (3) stealthiness of privacy leakage, and (4) necessity of shadow-model calibration.

\subsection{Privacy Leakage} 
\label{sec:eval-leakage} 
Fig.~\ref{fig:privacy-leakge-all-datasets} presents the attack ROC curves on different models. 

Both the uncorrupted models and the poisoned models by Tramer et al. exhibit different degrees of privacy leakage across settings. 
On the uncorrupted models, the attack TPR@0.1\% FPR varies from 0.76\% to 43.41\%, with an average 13.01\% TPR@0.1\% FPR. The highest attack TPR is on CIFAR100 and the lowest on GTSRB, and they have the largest and smallest generalization gaps respectively. 

The attack by Tramer et al. amplifies the privacy leakage, and increases the attack TPR from 13.06\% to 34.57\% (on CIFAR10), 43.41\% to 71.95\% (on CIFAR100), 6.85\% to 29.58\% (on SVHN), 0.76\% to 10.41\% (on GTSRB), and 2.09\% to 12.43\% (on PathMNIST). 
On average, this attack yields an attack TPR of 31.79\%. 

In comparison, \textbf{our attack consistently achieves high MI success with low false positives.} 
Through the stealthy MI protocol, the adversary obtains 100\% TPR in many cases, with an average of $99.99\%$ TPR@0.1\% FPR.  
Further, such high privacy leakage is achieved  across different architectures with various capacities and different training-set sizes (Section~\ref{sec:atk-analysis}).

\begin{figure}[!t]
  \centering
  \includegraphics[width=3.5in, height=1.in]{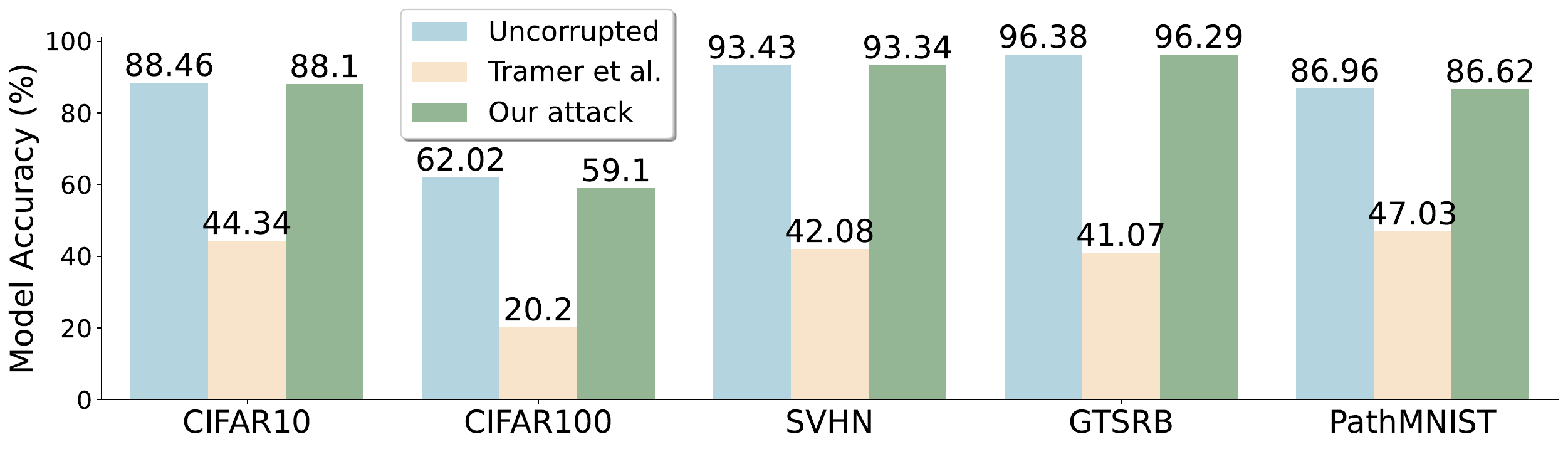}
  \caption{Model accuracy evaluation. The proposed attack consistently produces models with competitive accuracy.}
  \label{fig:acy-eval} 
  \vspace{-4mm}
\end{figure}

\subsection{Model Accuracy}
\label{sec:eval-utility}
Fig.~\ref{fig:acy-eval} reports the model accuracy. 
The attack by Tramer et al. incurs significant accuracy drop due to the injection of mis-labeled samples. 
It incurs  39.93\%$\sim$55.31\% accuracy drop, with an average of 46.51\%.

In comparison, \textbf{the poisoned models by our attack maintain comparable accuracy as the uncorrupted models}. 
The largest accuracy drop by our attack is 2.92\% on CIFAR100 (from 62.02\% to 59.1\%), which translates to a small 7.3\% increase of test error; and the average accuracy drop is only 0.77\%. 
{
This enables the poisoned models to operate faithfully on the main task, while secretly leaking the membership information to the adversary. 
}

\subsection{Stealthiness of Privacy Leakage}
\label{sec:eval-atk-stealthiness} 
Our attack succeeds in stealing the membership information through the proposed stealthy MI protocol, which is different from the standard MI protocol in existing work~\cite{shokri2017membership,yeom2018privacy,song2021systematic,choquette2021label,li2020membership,liu2022membership,wen2022canary,ye2021enhanced,carlini2022membership}. 
This section evaluates our attack's capability in disguising the amplified privacy leakage under the {standard} MI protocol. 

Under the standard MI protocol, for the attack by Tramer et al., a user without any knowledge of the data-poisoning adversary can still use the training samples to query the poisoned models and identify the high privacy leakage (average 31.79\% attack TPR@0.1\% FPR). 
This is undesirable from an adversary perspective as it can lead to a direct attack exposure. 

In contrast, \textbf{the poisoned models by our attack exhibit comparable privacy as their uncorrupted counterparts, under the standard MI protocol. } 
In Fig.~\ref{fig:privacy-leakge-all-datasets}, the attack TPRs between the code-poisoned models (\emph{green} lines) and uncorrupted models (\emph{blue} lines) are 12.46\% vs. 13.06\% (CIFAR10), 41.51\% vs. 42.30\% (CIFAR100), 6.75\% vs. 6.85\% (SVHN), 0.62\% vs. 0.76\% (GTSRB), and 3.18\% vs. 2.09\% (PathMNIST). 
The average TPRs are 13.01\%  and 12.91\%, respectively. 
Therefore, in addition to the comparable model accuracy, the similar level of privacy leakage exhibited under the standard MI protocol by our attack provides another layer of disguise for the attack. 

Moreover, our defense evaluation in Section~\ref{sec:defense-eval} shows our attack can be constructed to evade a state-of-the-art defense technique~\cite{tang2022mitigating} to exhibit strong privacy protection under the standard MI protocol.
This can tempt the users into believing that their models are ``private''; yet in reality, the adversary can continue to steal membership privacy in a secret manner. 

\begin{figure}[!t]
  \centering
  \includegraphics[ height=1.5in]{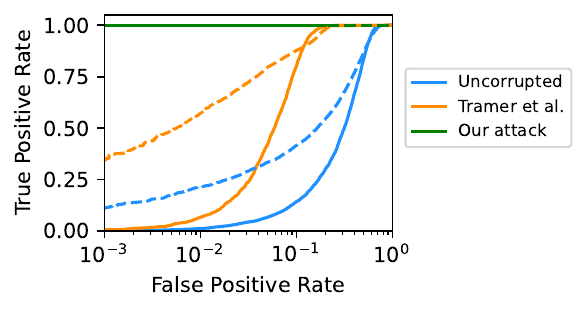}
  \caption{Comparing the membership inference success with (\emph{dashed} lines) and without (\emph{solid} lines) shadow-model calibration. 
  Without relying on shadow-model calibration, our attack can still de-identify all training members with low FPR.}
  \label{fig:other-mia-eval} 
  \vspace{-4mm}
\end{figure}

\subsection{Necessity of Shadow-model Calibration}
\label{sec:eval-atk-budget}
Training shadow models is commonly needed in existing attacks to calibrate the inference threshold in order to control at low FPR~\cite{carlini2022membership,wen2022canary,ye2021enhanced,tramer2022truth}. 
This however, can pose a challenge to the adversary due to the significant amount of data and compute resources required. 
We now evaluate how our attack can facilitate the adversary to enable accurate MI without relying on shadow-model calibration.  

We use the global-threshold-based variant in LiRA~\cite{carlini2022membership}, which does not
require shadow models to perform the fine-grained per-sample calibration. 
We report the results on CIFAR10 in Fig.~\ref{fig:other-mia-eval} (and we observe similar trends on other datasets and {on using other generic attacks such as the loss- and confidence-based attack~\cite{yeom2018privacy}}). 

On the uncorrupted model, the global-threshold attack fails to infer the member samples when controlled at low false positive. 
Even when the model was trained with poisoned data \cite{tramer2022truth}, this attack is still unsuccessful  
and it can only achieve 0.55\% attack TPR@0.1\% FPR (the orange solid line in Fig.~\ref{fig:other-mia-eval}). 
For the adversary to expose the amplified privacy leakage, he/she still has to resort to shadow-model calibration, which achieves 34.57\% TPR@0.1\% FPR. 

In comparison, \textbf{our attack succeeds in performing accurate MI without shadow-model calibration}. 
On the code-poisoned models, the global attack achieves the same 100\%  TPR@0.1\% FPR as when using shadow-model calibration.
This renders our attack much more practical than prior work.

\subsection{Additional Analysis} 
\label{sec:atk-analysis}
This section conducts further analysis into the proposed attack using CIFAR10. 
Section~\ref{sec:diff-models} evaluates our attack under \textbf{models with various capacities}, while   Appendix~\ref{sec:diff-sizes} shows the evaluation on \textbf{different training-set sizes}.  
Section~\ref{sec:comparison-song-et-al} \textbf{compares the basic attack with the complete attack} on all evaluation settings. 
We evaluate our attack under \textbf{different defense techniques} in Section~\ref{sec:defense-eval}, and finally present an \textbf{ablation study} in Appendix~\ref{sec:ablation}. 
For the poisoned models, we follow the stealthy MI protocol and use the generic attacks without shadow-model calibration, which has the benefit of saving the cost in training shadow models.

\begin{table}[t]
\centering 
\normalsize
\renewcommand{\arraystretch}{1.1}
\caption{Evaluating the proposed attack on models with different capacities. Accuracy drop measures the accuracy difference between the poisoned and uncorrupted models. }
\label{tab:eval-diff-arch}
\begin{tabular*}{\columnwidth}{@{\extracolsep{\fill}}ll|ccc }
\hline
\multirow{2}{4em}{Architecture} & \multirow{2}{4em}{Parameter size \#} & \multirow{2}{3em}{Poison acy.} & \multirow{2}{2em}{Acy. drop} & {TPR@} \\
& & & & 0.1\% FPR \\
\hline    
WRN-28-10& 36.51M  & 88.10 & -0.36 & 100.00 \\
WRN-28-7 & 17.89M  & 87.32 & -1.03 &  99.98 \\
AlexNet & 14.86M & 78.57 & -1.54 & 98.91 \\
SENet-18 & 11.27M & 86.80 & -0.52 & 100.00 \\
ResNet-18& 11.18M & 85.50 & -0.18 & 99.98\\
WRN-16-8 & 10.97M & 86.43 & -0.75 & 100.00\\
ResNeXt & 9.15M & 85.50 & -1.24 & 99.98 \\
WRN-40-4 & 8.97M  & 86.75 & -1.36 & 99.86 \\
DenseNet-121&7.04M & 87.30 & -0.64 & 99.09\\
GoogleNet & 6.18M & 86.68 & -0.77 & 100.00 \\
WRN-28-4 & 5.86M  & 87.39 & -0.20 & 100.00\\
WRN-40-2 & 2.25M  & 85.83 & -1.40 & 99.43 \\
WRN-28-2 & 1.47M  & 85.61 & -1.03 &  99.10 \\
\hline
Average & & & -0.85 & 99.72 \\
\hline
\end{tabular*}
\end{table}

\subsubsection{\textbf{Evaluation on models with different capacities}}
\label{sec:diff-models}
Our attack amplifies the membership leakage through manipulating the model to memorize the membership-encoding samples. 
While deep learning models are 
capable of memorizing data~\cite{zhang2021understanding,feldman2020neural,feldman2020does}, 
it is known that the memorization effect is related to model capacity. 
We thus evaluate our attack under models with different capacities. 

We vary the layer depth and the widening  factor in the WideResNet architecture and also consider six other network architectures: DenseNet~\cite{huang2017densely}, SENet~\cite{hu2018squeeze} ResNeXt~\cite{xie2017aggregated}, {ResNet~\cite{he2016deep}, AlexNet~\cite{krizhevsky2012imagenet} and GoogleNet~\cite{szegedy2015going}}, totaling 13 different models with diverse parameter sizes. 
Table~\ref{tab:eval-diff-arch} reports the results. 
Our attack maintains the high success even when the model capacity is reduced by 25x (from 36.51M to 1.47M). 
On average, our attack achieves 99.72\% TPR@0.1\% FPR and 0.85\% accuracy drop.   

Aggressive truncation of model capacity (e.g., from 36.51M to 0.57M) can reduce attack TPR to 45\%. 
However, due to the limited model capacity, even the uncorrupted model (trained without our attack) cannot obtain high accuracy, and it increases the test error by > 23\% compared with that on the larger models (undesirable).

\subsubsection{{\textbf{The basic attack Vs. the complete attack}}}
\label{sec:comparison-song-et-al} 
The basic attack in Section~\ref{sec:basic-atk} represents an extension of the reconstruction attack by Song et al.~\cite{song2017machine}, and we compare it with our complete attack approach. 
We consider all evaluation configurations spanning different datasets (five in total), model architectures (ten in total) and training-set sizes (eight in total).  

\begin{figure}[!t]
  \centering
  \includegraphics[ height=1.5in]{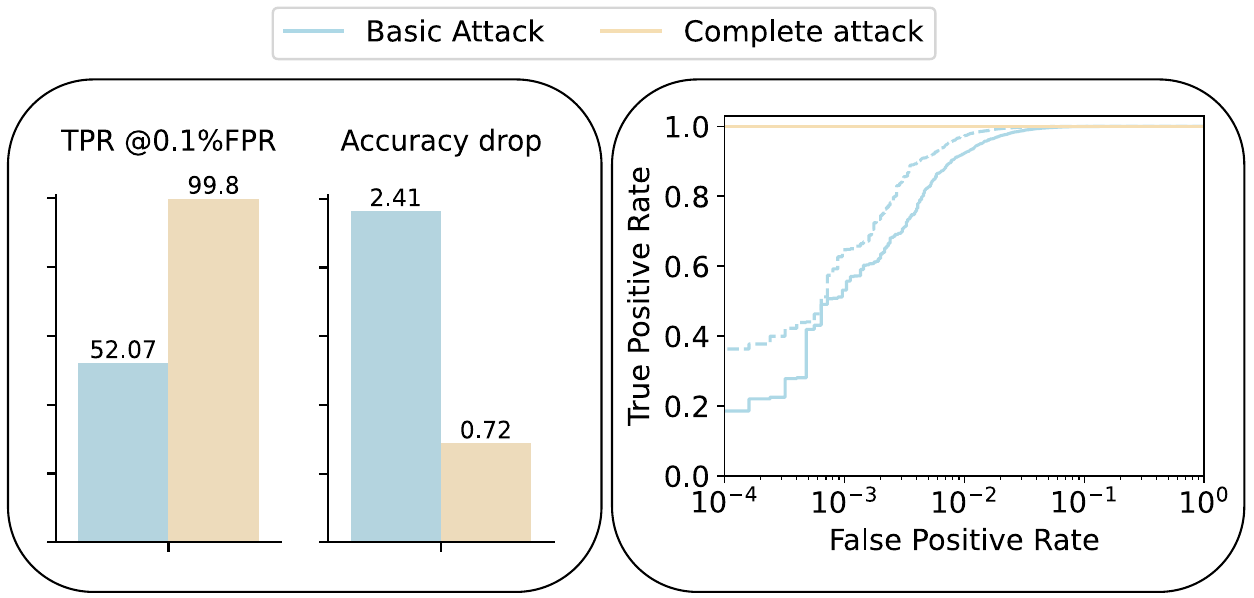}
  \caption{Comparing the basic attack with the complete attack approach. \emph{Left}: Average results across all evaluation settings (datasets, models, training-set sizes). \emph{Right}: Example from CIFAR10, where the complete attack maintains superior advantage even if the basic attack is calibrated with shadow models (the blue dashed line). Overall, the complete attack approach achieves considerably better attack performance (99.80\% TPR@0.1\% FPR vs. 52.07\% TPR) and much lower accuracy drop (70\% lower).}
  \label{fig:song-et-al} 
  \vspace{-2mm}
\end{figure}

As shown in  Fig.~\ref{fig:song-et-al}, the complete attack achieves significantly higher MI success when controlled at low FPR regime (99.8\% TPR@0.1\%FPR vs. 52.07\% by the basic attack), and also with much lower accuracy loss (70\% lower). 
In addition, we also evaluate the attack performance when the basic attack is calibrated with shadow models (the blue dashed line on the right of Fig.~\ref{fig:song-et-al}), and find that the complete attack still maintains superior advantage. 

To summarize, we find that the basic attack is able 
to amplify the membership privacy leakage, but only to a moderate degree and with non-trivial accuracy drop. 
This is due to a problem we identify as distribution mismatch.  
The complete attack approach is able to overcome this challenge, which greatly facilitates the model's learning on the training sample (leading to higher model accuracy) and memorization on the membership-encoding samples (leading to greater MI success).

\begin{table}[t]
\centering 
\footnotesize
\caption{Summarizing existing defense techniques and their performance characteristics. In generic regularization techniques, strong regularization leads to high privacy but low accuracy (vice versa). 
Our attack can be constructed to evade those defenses based on their defensive loss terms, via optimizing both the defensive and malicious loss terms. 
}
\label{tab:mia-summary}
\begin{tabular*}{\columnwidth}{@{\extracolsep{\fill}}l|ccc}
\hline
\multirow{2}{6em}{Defense type} & Defensive & Privacy  & Accuracy   \\
                            & loss? & protection & drop\\
\hline
Provable defense~\cite{abadi2016deep,papernot2016semi} & \xmark & Strong & High   \\
Soft-label based~\cite{tang2022mitigating,shejwalkar2019membership,chen2023overconfidence} & \cmark & Weak & Low   \\
Add training constraint~\cite{li2021membership,chen2023overconfidence} & \cmark & Weak & Low \\
Output obfuscation~\cite{jia2019memguard,chen2023overconfidence} & \xmark & Weak & Low  \\
Generic regularization~\cite{yao2007early,srivastava2014dropout} & \xmark & Weak/Strong & Low/High \\
\hline
\end{tabular*}
\end{table} 

\subsubsection{\textbf{Defense evaluation}}
\label{sec:defense-eval}
Recall that our attack modifies the loss-value function, and hence depending on whether the defense has its defensive loss term, our attack may be constructed to evade it. 
Table~\ref{tab:mia-summary} categorizes existing defenses based on whether they have their own defensive loss terms.

\emph{Evade-able defenses}. For those defenses that consist of their own defensive loss terms (e.g., the KL loss for soft label training in \cite{shejwalkar2019membership,tang2022mitigating,chen2023overconfidence}), the adversary can create a malicious loss term for the membership-encoding samples and optimize the model on both losses. 
Here, our goal is to \emph{evade} the defense and produce compromised models that can exhibit strong privacy under the standard MI evaluation. This would  mislead the users into believing that their models are ``private'', yet in reality the models still allow the adversary to perform accurate MIA. 
It can conceal our attack within a seemingly private model, and render the attack 
even more insidious. 

\emph{Non evade-able defenses}. On the other hand, there exist other defenses that cannot be evaded by the malicious loss (e.g., DPSGD~\cite{abadi2016deep} that performs gradient perturbation). 
In this case,  we evaluate how these defenses can reduce the privacy leakage by our attack. 

We next discuss our evaluation on different defenses.

\begin{figure}[!t]
  \centering
  \includegraphics[width=3.4in, height=1.6in]{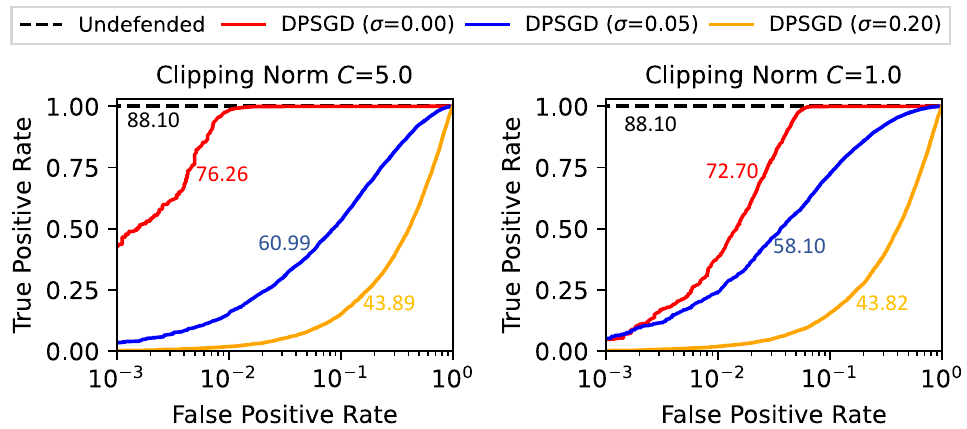}
  \caption{Evaluating DPSGD under different clipping norms ($C$) and noise multipliers ($\sigma$). The numbers along each curve give the model accuracy. 
  Overall, DPSGD is an effective defense against our attack, though it also incurs high accuracy loss. 
  The more the noise injected, the stronger is the privacy protection, and the lower is the model accuracy.}
  \label{fig:dpsgd} 
  \vspace{-4mm}
\end{figure}

\textbf{Provable defense. } 
We consider DPSGD, a principled defense based on differential privacy~\cite{dwork2006calibrating,abadi2016deep}.
It bounds the influence that any sample can have on the model via performing clipping and noise injection to the gradients derived from {all} loss terms (hence it cannot be evaded by our attack). 
We evaluate different clipping norms $C\in \{1, 5\}$ and noise multipliers $\sigma\in \{0.0, 0.05, 0.2\}$, and  Fig.~\ref{fig:dpsgd} reports the results.

Using a tight clipping norm without injecting any noise (the red curve on the right of Fig.~\ref{fig:dpsgd}), DPSGD  reduces the attack TPR@0.1\% FPR to only 4.74\%. However, it also causes a high accuracy drop of 14\%.  
We also find that under DPSGD, our attack causes additional accuracy loss (average 3.4\%), compared with the models trained without poisoning. 
Overall, injecting more noise further improves the privacy, but also incurs higher accuracy loss. 

The trade off between privacy protection and model accuracy has been a longstanding challenge, and there are many defenses that aim to preserve high model accuracy while providing strong empirical privacy protection. 
We next discuss our evaluation on several such defenses. 

\textbf{Soft-label based defense (\emph{evade-able}).}
We  consider the state-of-the-art SELENA defense (USENIX Security'22)~\cite{tang2022mitigating}, based on knowledge distillation. We explain it next.

First, SELENA partitions the training set into different subsets, each of which is used to train a teacher model. 
For each subset, there exists another set of remaining samples that are not used to train the corresponding teacher model, and these samples are viewed  as the reference {non-member} samples. 
Then, the teacher models make predictions on their respective reference samples,  the outputs of which are aggregated as the {privacy-preserving soft labels} to be learned by the student model. 
In essence, the student model is trained to predict the sensitive training samples as if they are the non-member samples, which can mitigate the model's overfitting. 

\begin{figure}[!t]
  \centering
  \includegraphics[ height=1.6in]{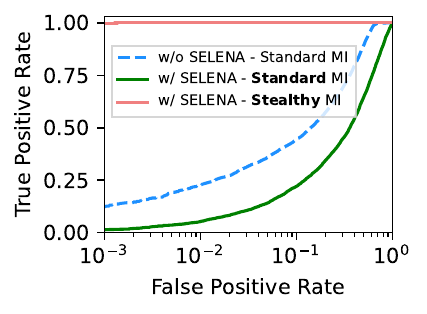}
    \caption{
    Under the standard MI protocol, the model trained with the SELENA defense exhibits strong privacy protection and reduces the attack TPR@0.1\% FPR from 12.46\% (blue dashed line) to  1.72\% (green solid line). 
    However, our attack can be constructed to evade the defense and maintain high MI success (red line). 
    }
  \label{fig:selena} 
  \vspace{-4mm}
\end{figure}

We first explain how we construct our attack while incorporating the SELENA defense into the training code. 
The privacy protection in SELENA relies on a set of privacy-preserving soft labels for performing knowledge distillation. 
However, they are only applied to the original training samples to derive a \emph{defensive} loss term, and our attack can create another \emph{malicious} loss value to evade the defense, by using both losses to optimize the model. 
As the adversary can use {arbitrary} labels for the membership-encoding samples, he/she can generate soft labels whose top-1 classes have 99\% probability, for the membership-encoding samples. 
This produces a malicious loss that significantly facilitates the model's memorization on the synthetic samples and contributes to the high attack success.

Fig.~\ref{fig:selena} shows the results. 
Under the standard MI protocol, the model exhibits strong protection (offered by the SELENA defense) and reduces the attack TPR from 12.46\% to only 1.72\%, with a small  accuracy drop of 3.35\% (similar accuracy drop on the model trained without our attack). 
Under the stealthy MI protocol, however, the adversary can still achieve a very high 99.73\% TPR. 

In the above, we demonstrate that \emph{the proposed attack can be  constructed to produce a poisoned model that conveys a  sense of ``strong'' privacy under the standard MI protocol, yet it still secretly leaks the membership information to the adversary}. 
This can mislead the users into believing that the compromised models they have are ``private'', and render our attack even more insidious. 
For the adversary, he/she can provide the privacy-preserving training algorithm as an option in the training code. 
The users can decide to train the model with the standard (without defense) or defensive (with defense) loss term - 
in \emph{either} case, our attack can continue to inflict significant privacy leakage via the compromised loss. 

The use of soft labels 
in SELENA is a common approach and is also employed in other related defenses such as DMP~\cite{shejwalkar2021membership} and Label Smoothing~\cite{szegedy2016rethinking}. 
Thus, the success of our attack in evading the SELENA defense also has repercussions on these defenses, e.g., the adversary can provide these techniques as different training options in the training code for the users to choose from.

\textbf{Defense based on adding training constraint  (\emph{evade-able}).}  
This class of defense adds optimization constraints during training to regularize the model's behavior and reduce its privacy risk~\cite{li2021membership,chen2023overconfidence,nasr2018machine}. 
We evaluate a representative technique based on regularizing the output distributions~\cite{li2021membership}, and show that it can be similarly evaded by optimizing the defensive and malicious loss term together (Appendix~\ref{sec:eval-regularization}). 

\textbf{Output perturbation. } 
This line of defense performs output obfuscation on the trained model~\cite{jia2019memguard,chen2023overconfidence} and hence our attack cannot evade them. 
We evaluate two representative defenses~\cite{jia2019memguard,chen2023overconfidence}, and find that our attack still achieves high success despite the obfuscated outputs (Appendix~\ref{sec:eval-outputMod}).

\textbf{Generic regularization.} 
Generic techniques such as early stopping~\cite{yao2007early}, dropout~\cite{srivastava2014dropout} are helpful in mitigating model overfitting and reducing privacy risk. 
However, they generally suffer from the trade off between privacy protection and model accuracy~\cite{salem2018ml,shejwalkar2019membership}, and we validate this in Appendix~\ref{sec:eval-generic-reg}. 

%% file: tex/discussion.tex
\label{sec:discussion}

We first evaluate our attack under the normalization-layer-free setting (Section~\ref{sec:norm-free}), and then perform a comprehensive study to understand the artifacts incurred by our attack (Section~\ref{sec:attack-artifact}). 
Section~\ref{sec:countermeasure} presents an attack countermeasure and Section~\ref{sec:limitation} discusses the limitations of our work. 

\subsection{Is Normalization Layer Indispensable?} 
\label{sec:norm-free}
The proposed architectural modification to include a secondary normalization layer plays a key role in our attack. 
While the norm layer is a common building block in many state-of-the-art deep learning models~\cite{he2016deep,huang2017densely,zagoruyko2016wide,hu2018squeeze,xie2017aggregated}, there are models that do not include the norm layer
~\cite{brock2021characterizing,brock2021high}. 
This section analyzes our attack under such a normalization-free model. 

We use a simpler approach by removing all the norm layers in the WideResNet model. 
Without the norm layer to shift and scale the inputs, we switch to using a larger mean of 0.3 and standard deviation of 1.5 for generating diverse membership-encoding samples (otherwise they cannot be memorized due to the low variance across samples).
In this setting, our attack performance is not as high, though it still increases the TPR@0.1\% FPR from 9.96\% to 48.29\% (a 4.8x increase) with 0.9\% accuracy drop. 
We leave the improvement under the normalization-free setting to future work.

\subsection{{Discussion on Attack Artifacts}}
\label{sec:attack-artifact} 
In this section, we first discuss the different artifacts incurred by our attack, and then present several alternative strategies to mitigate them. 

\textcircled{\normalsize 1} {Additional forward passes}. 
Our attack creates membership-encoding samples during training. 
They are used to create the malicious loss value, but also increase the number of forward passes at each training step. 
Therefore, we present an alternate method to configure our attack without increasing the number of forward passes. 

The idea is to randomly \emph{replace} a subset of training samples with their membership-encoding samples, while keeping the number of forward passes the same at each training step (e.g., 70\% for training samples, and 30\% for synthetic samples). 
The rationale is that both types of samples do not need to appear in every training step to be learned/memorized by the model, and thus we can steal the membership without increasing the number of forward passes, which also has the benefit of reducing the attack overhead (see \textcircled{\normalsize 2} next). 

We evaluate our attack by replacing different portions of training samples with membership-encoding samples at each training step (10\%, 30\%, 50\% and 70\%). 
10\% replacement can largely preserve the model accuracy (0.65\% drop), but there are limited number of  membership-encoding samples at each step, which leads to a slightly lower TPR of 80\%. 
Increasing the ratio to 30\% can boost the TPR to 99.82\%, at a slightly higher accuracy drop of 1.1\%. 
Using a larger replacement ratio has negligible benefit in privacy leakage and incurs slightly higher accuracy loss (as fewer training samples are used at each training step):  50\% replacement ratio has a 1.5\% accuracy drop and 70\% has 3.44\%\footnote{The replacement is done randomly and thus all training samples will still be seen by the model during training to obtain high prediction accuracy.}.

\textcircled{\normalsize 2} {Increased model complexity and overhead}. 
The inclusion of a secondary norm layer increases the model parameters, but only to a small margin (average 0.15\%). 
We also measure the runtime overhead by our attack (with two Nvidia V100 GPUs). 
The attack increases the average training time (from five repetitions) from $85.4$ mins to $165$ mins (93\% increase).  However, using the random replacement strategy introduced earlier (with a 30\% replacement ratio) can reduce the runtime to 108.2 mins, which amounts to a mere 26.7\% increase. 
Lastly, our attack also increases the average inference overhead from $7.51$ ms to $7.87$ ms (4.8\% higher).  

Despite the increased overhead, we remark that they do {not} make the attack easy to detect.  This is because 
the runtime is also affected by several other factors (e.g., system environment, hardware configuration), and hence it is challenging to obtain stable runtime baselines for comparison.

\textcircled{\normalsize 3} {The secondary normalization layer leaves an artifact in the model's computational graph}. 
{
However, to the best of our knowledge, the \emph{existing} use of additional normalization layer in ML models is commonly intended for \emph{benign} purposes (such as for adversarial training~\cite{xie2020adversarial,xie2019intriguing}, or performance improvement under imbalanced classification~\cite{zada2022pure}). 
In contrast, our work is the \emph{first} of its type to exploit the additional norm layer from an \emph{adversarial} perspective to facilitate MIAs. 
Therefore, it is highly challenging for non-expert target users to determine that the extra norm layer was included for a malicious intent. 


}

{
For completeness, we also investigate other alternative attack strategies that do \emph{not} entail architectural modification (details in Appendix~\ref{sec:mgda-alternative}). 
This can be adopted by the adversary to eliminate the artifact of the secondary norm layer while still causing major privacy damage, though they do come with the cost of reduced attack performance (Appendix~\ref{sec:mgda-alternative}). 
}

{
\emph{Summary}. 
We comprehensively discussed the artifacts incurred by our attack, and also explained carefully why these artifacts do not make our attack easy to detect. 
In addition, we also presented several {alternative methods} that can greatly mitigate the attack artifacts 
while still inflicting considerable privacy leakage. 

Overall, our attack represents a new class of MIA with several noteworthy properties, including: (1) high MI success against all training samples (average $>$99\% attack TPR@0.1\% FPR), (2) no reliance on shadow model calibration, and more importantly, (3) incurring negligible accuracy drop, while (4) being able to disguise the amplified privacy leakage under common membership privacy auditing. 
These together represent a significant advancement over existing poisoning-based MIAs~\cite{song2019robust,chen2022amplifying,tramer2022truth}. We leave further improvement in covering the remaining attack traces to follow-up studies.  
}

\subsection{Attack Countermeasure.}
\label{sec:countermeasure}
Our evaluation in Section~\ref{sec:defense-eval} on various MIAs defenses illustrates the challenges in mitigating the proposed attack. However, our attack relies on the exact knowledge of the target sample, for generating the unique random seed to reconstruct the membership-encoding sample. 
Hence, one countermeasure is to slightly modify the target sample so that the adversary cannot generate the same random seed to reconstruct the secret sample. 
The modifications can manifest in different  forms, and thus they are hard to predict by the adversary. 

With that said, the exact knowledge of target sample is still a common assumption in existing practice of MIAs~\cite{shokri2017membership,ye2021enhanced,carlini2022membership,hu2022membershipsurvey}. Hence, our attack still poses significant privacy threat, and a systematic amendment of the existing MI defenses to handle our attack is another avenue for future work.

\subsection{{Limitation}}
\label{sec:limitation}
The major limitation of our work concerns the {feasibility of mounting code poisoning attacks in practice}. 
While code poisoning attacks have been shown to be feasible in real-world ML codebase~\cite{pytorch-code-poisoning,pytorch-code-poisoning1,code-poisoning2}, 
we have not realized  our attack in the open world, and neither have {any} prior code-poisoning attack studies~\cite{bagdasaryan2021blind,song2017machine,song2019robust,liu2022loneneuron}, to the best of our knowledge. This is a limitation of this class of studies.

Nevertheless, launching the attack in the open world 
would require strong controls to prevent any actual harm   
to users of the ML infrastructure. 
Any {controlled} attack that seeks to amplify the privacy leakage of production ML models should carefully consider how to address the related ethical concerns, an open question that requires further research. 

{
Finally, while our attack is yet to be realized in practice, we hope the comprehensive proof of concept of the proposed attack can bring awareness to the risk of code poisoning in third-party ML codebase, which becomes highly relevant given that integrating external code repositories in the development of ML model is becoming a growing practice. 
Thus, our work also calls for future efforts on ML code security and we outline several such directions next. 
}

%% file: tex/conclusion.tex
\label{sec:conclusion}
This work introduces a new form of membership inference attack against deep learning models, based on poisoning two opaque and difficult-to-test modules in the model-training code: the loss-value computation and model structure. 
The training code can be used by the victim users in a trusted environment to produce compromised models that can operate faithfully on the main task with competitive performance, while still secretly leaking the membership information of all the training samples to the black-box adversary. 

Our work illustrates how the massive learning capacity of modern deep learning models can be exploited by the adversary to amplify membership privacy leakage in a secret manner. The amplified privacy leakage inflicted by the attack can remain \emph{unnoticeable} under common privacy auditing methods,  and a deliberate  adversary can go even further to disguise the attack by tricking the corrupted model to convey a \emph{false} sense of strong privacy and mislead the users. 
From this, we outline three directions to be explored in future studies. 

\textbf{(1) Rethinking the current membership privacy auditing practice}. 
Existing auditing practice does not account for code inspection, which is a necessary step in exposing the privacy leakage inflicted by our attack. 
Thus, an open question is should the model-training code be supplied as part of the inputs in the standard membership inference game, though identifying the malicious constructs from the complicated codebase itself can become {another} barrier? 
In addition, a direct approach to thwart our current attack is to slightly modify the target sample.  Thus, developing a standardized approach to enact this can be another avenue for future study. 

\textbf{(2) Extending our attack to other settings.}
There are several directions that can be explored to extend our attack, 
including attack extension to generative models~\cite{hayes2017logan,chen2020gan}, to other domains such as natural language processing~\cite{shejwalkar2021membership,mireshghallah2022quantifying},  
and improving the attack performance under other challenging settings (e.g., normalization-free scenario). 

\textbf{(3) Developing more capable defenses and code analysis tools.}
Our evaluation on existing defense techniques shows that the challenge in providing  strong privacy protection without incurring high accuracy loss still remains. 
Those prior privacy defenses that can achieve a superior privacy-utility trade off under standard MI evaluation, unfortunately can be ``evaded'' by a deliberate adversary, and 
future work can study the potential of more capable defenses to withstand such attack. 

Another related direction is to develop automated code inspection tools to analyze the irregular and potentially malicious logic in the ML codebase. 
There are a number of code integrity checking tools for traditional software~\cite{burow2017control,castro2006securing}.  Future work can study whether they can be adapted to counter code poisoning attacks in ML codebases, where the key challenge is that a similarly irregular code logic can be intended for either a benign or malicious purpose. 
{
For example, a trusted computational graph~\cite{bagdasaryan2021blind} can be used to detect whether the malicious code has caused the model to deviate from the expected computational graph during training, such as creating an additional forward pass. 
However, the adversary can still replace a subset of training samples with the membership-encoding samples to mount the attack without causing an extra forward pass (Section~\ref{sec:attack-artifact}). 
} 
Software signing~\cite{signing-1, signing-2,newman2022sigstore}, where the maintainers  digitally sign their released codebases to prevent unauthorized code manipulation, is also a promising direction. 
However, in modern ML ecosystems where many codebases are maintained by multiple developers and undergo frequent iterations, the potential usability concerns~\cite{newman2022sigstore,schorlemmer2024signing} are another factor to be considered.

%% file: tex/appendix.tex
\subsection{Evaluation on Different Training Sizes}
\label{sec:diff-sizes}

This section evaluates our attack under different configurations of training-set size. The results are shown in Table~\ref{tab:eval-diff-size}. 

In terms of model accuracy, the poisoned models consistently maintain comparable accuracy as the uncorrupted counterparts, with an average accuracy drop of 0.43\%. 

In terms of privacy leakage, the proposed attack also succeeds under a wide range of training sizes, and achieves an average of $>$99\% TPR@0.1\% FPR across different sizes.

\begin{table}[!t]
\centering 
\normalsize
\caption{Attack evaluation on different training-set sizes. 
{Our attack consistently achieves high privacy leakage with low accuracy drop}.}
\label{tab:eval-diff-size}
\renewcommand{\arraystretch}{1.1}
\begin{tabular}{llll }
\hline
\multirow{2}{6em}{Training-set size} & \multirow{2}{5em}{Model accuracy} & \multirow{2}{5em}{Accuracy drop} & Attack TPR \\
& &  & @0.1\% FPR \\
\hline                                                      
2,500 & 69.38 & -0.03 & 99.80 \\
5,000 & 79.27 & -0.56 & 99.24 \\
7,500 & 83.32  & -0.17 & 99.97 \\
10,000& 84.71 & -0.81 & 99.65 \\
12,500& 88.10 & -0.30 & 100.00 \\
15,000& 89.06  & -0.47 & 100.00 \\
20,000& 90.25  & -0.81 & 99.85 \\
25,000& 92.27  & -0.22 & 100.00 \\
\hline
Average &  & -0.43 & 99.81 \\
\hline
\end{tabular}
\end{table}

\subsection{Additional Results on Defense Evaluation}
\label{sec:res-more-defenses} 
This section reports the results on three different types of defenses based on: (1) adding training constraint, (2) output perturbation, and (3) generic regularization techniques.

\subsubsection{\textbf{Defense based on adding training constraint}}
\label{sec:eval-regularization}
We evaluate the defensive regularization based on Maximum Mean Discrepancy (MMD) in Li et al.~\cite{li2021membership}. 
The idea is to regularize the MMD distance between the softmax output distributions of the training members and the validation (non-member) samples, which essentially encourages the model to predict the training samples as if they are the  non-members. 

\begin{figure}[!t]
  \centering
  \includegraphics[ height=1.5in]{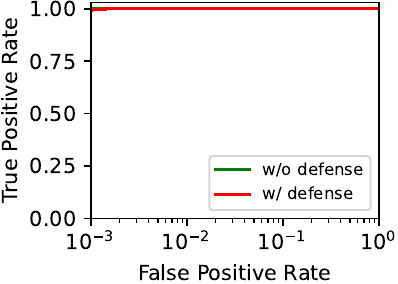}
    \caption{Evaluation on the defense by adding optimization constraint~\cite{li2021membership}. Our attack can be constructed to evade the defense and maintain its high success (both lines overlap).}
  \label{fig:mmd-eval} 
  \vspace{-3mm}
\end{figure}

We evaluate this defense using different five different parameters from $[1, 7]$ to vary the regularization strength (larger values cause excessive regularization, and the model cannot be trained to obtain good accuracy). 
We show the results in Fig.~\ref{fig:mmd-eval} on one setting - our attack achieves consistently high success in all settings ($>$99\% TPR@0.1\% FPR and average $<$1\% accuracy drop). 
This is because the regularization term is applied only to the outputs on the original training samples, and our attack can create the malicious loss term from the membership-encoding samples to maintain its success (similar to what we did on the SELENA defense in Section~\ref{sec:defense-eval}).  

There are other related defenses with different regularization forms such as the min-max game in AdvReg~\cite{nasr2018machine}, and our attack can be constructed in a similar way to evade them.

\subsubsection{\textbf{Defense based on output obfuscation} }
\label{sec:eval-outputMod}

These defenses seek to obfuscate the output vector to reduce privacy leakage. Representative techniques include MemGuard, which perturbs the output vector to confuse a shadow MI classifier~\cite{jia2019memguard}; and HAMP\footnote{{HAMP consists of soft label training, adding training constraint, and output obfuscation. Our earlier evaluations show that the first two components can be evaded by our attack (as in Section~\ref{sec:defense-eval} and Appendix~\ref{sec:eval-regularization}), and hence we evaluate the last component (output obfuscation defense) against our attack.}}, which replaces the output vector with a randomized vector (from the output on a randomly created sample) while preserving only the relative ordering within the vector, i.e., only the label-related information are preserved~\cite{chen2023overconfidence}.  

\begin{figure}[htb]
  \centering
  \includegraphics[ height=1.5in]{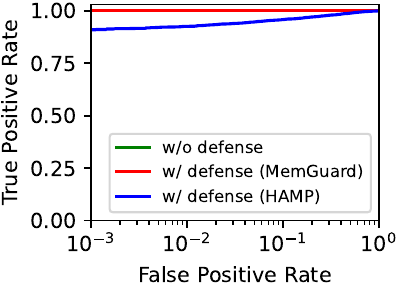}
    \caption{Evaluation on two defenses that perform output obfuscation (MemGuard~\cite{jia2019memguard} and HAMP~\cite{chen2023overconfidence}). Our attack continues to achieve very high MI success despite the obfuscated outputs (the green and red line overlap).}
  \label{fig:outputMod-eval} 
\end{figure}

We evaluate both defenses and report the results in Fig.~\ref{fig:outputMod-eval}.

For MemGuard, 
we first validate that it is able to reduce the attack accuracy of the shadow MI classifier from $\sim$99\% to 50\% (equivalent to random guessing). Yet prior work find that this approach of MemGuard still provides very limited privacy protection~\cite{tang2022mitigating,song2021systematic,shejwalkar2019membership},  and our result in Fig.~\ref{fig:outputMod-eval} (the red line) has a similar observation. 

{On the other hand, there are other approaches for obfuscating the output vectors. One representative technique is HAMP, which completely randomizes the output vectors~\cite{chen2023overconfidence}. }
In this case, attacks that compute a generic metric (e.g., prediction loss) are no longer effective, and thus we need another attack to retrieve the stolen membership. 
Since the adversary is not restricted to types of attacks he/she can use, we use the neural network (NN) based attack by Nasr et al.~\cite{nasr2018comprehensive}, which we find to be  effective against such randomization-based defenses. 

This attack trains a neural network to discern the model's outputs on members and non-members. This attack uses the ground-truth label, prediction loss, and the logit values as the input to train the model and outputs a membership probability. 
We follow prior work~\cite{nasr2018machine,tang2022mitigating,shejwalkar2019membership,chen2023overconfidence} to use the first half of the members and non-members to train the attack model, and perform evaluation on the remaining half. 
As in Fig.~\ref{fig:outputMod-eval}, even if the outputs are randomized to preserve only the label-related information, our attack still achieves very high MI success, with 90.98\% TPR@0.1\% FPR. 

Therefore, we find that defenses that obfuscate the outputs cannot provide meaningful protection against our attack.

\subsubsection{\textbf{Generic regularization techniques.}}
\label{sec:eval-generic-reg}
We evaluate two common techniques: dropout~\cite{srivastava2014dropout} and early stopping~\cite{yao2007early}.

\begin{figure}[t]
  \centering
  \includegraphics[ height=1.5in]{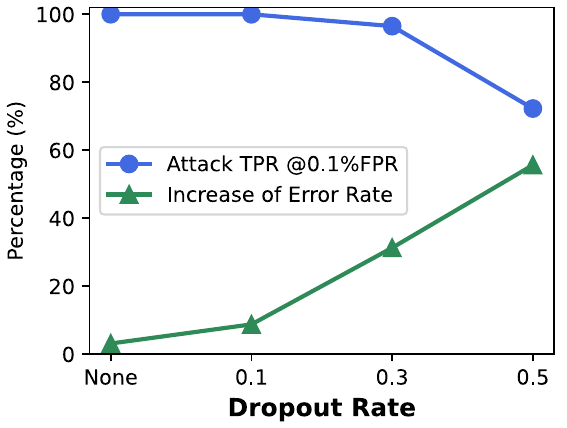}
    \caption{Evaluating dropout as a defense.  
    Our attack is resilient to moderate dropout; and aggressive dropout can partially mitigate the attack success, 
    but with severe accuracy drop.}
  \label{fig:dropout} 
  \vspace{-3mm}
\end{figure}

\emph{Dropout}. 
We report the results with different dropout rates in Fig.~\ref{fig:dropout}. 
Our attack is resilient when evaluated under a small dropout rate of 0.1 - in this setting, our attack can still achieve 99.99\% TPR@0.1\% FPR with 1\% accuracy loss. 
Using a larger dropout rate of 0.5 can partially foil  our attack to obtain a lower TPR of 72.26\%. 
However, under such an aggressive dropout, even the uncorrupted model (trained without any code poisoning) suffers from severe performance degradation, and the test error is increased by 55.8\% (similar increase of 55.6\% on the poisoned model). 
This indicates that using dropout as a defense technique suffers from the undesirable trade off between privacy protection and model accuracy. 

\begin{figure}[t]
  \centering
  \includegraphics[ height=1.5in]{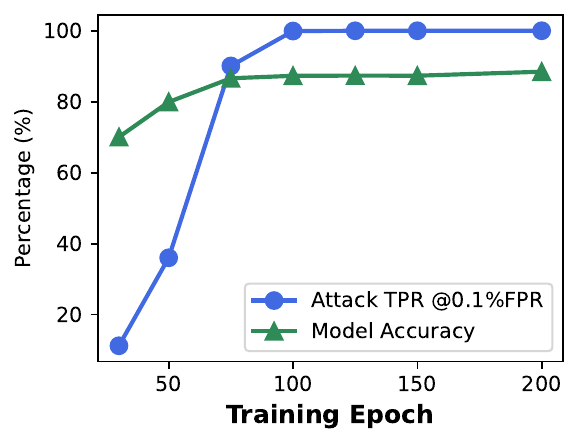}
    \caption{Evaluating early stopping as a defense. Our attack has low success when the model is in its underfitting stage, but the model also has low accuracy. 
    Increasing the training epochs increases the model accuracy, but so does the attack success.}
  \label{fig:earlyStop} 
  \vspace{-4mm}
\end{figure}

\emph{Early stopping}  is a generic regularization technique for reducing model overfit, and is known to be helpful in mitigating MIAs. 
To perform the evaluation, we benchmark the models trained in different epochs (under early stopping), and evaluate their accuracy and privacy leakage respectively. 

The results are in Fig.~\ref{fig:earlyStop}. 
We first see that our attack is less effective during the underfitting stage, which is understandable as the model is unlikely to exhibit any memorization at such an early stage. 
When trained with only 30 epochs, the model obtains 70.05\% accuracy (Vs. 88.1\% by full training), on which our attack achieves 11.24\% TPR@0.1\% FPR. 

After the early training stage, the attack performance improves rapidly. 
At epoch 50, the model obtains 79.92\% accuracy, and 
our attack achieves 36\% TPR. 
At epoch 75, our attack approaches its maximal performance and already achieves $>$90\% TPR (the model has still not converged as its accuracy of 86.61\% is still lower than the 88.1\% full accuracy). 
Our attack yields 99.92\% TPR after 100 epochs, on which the model has 87.3\% accuracy.

\subsection{\textbf{Ablation Study}}
\label{sec:ablation} 

{
Our attack consists of the following components: (1) mean and standard deviation for generating  membership-encoding samples, (2) the labels of these samples, and (3) the inclusion of secondary normalization layer. 
We next conduct an ablation study for each of these components. 
}

\subsubsection{\textbf{Impact of different mean and standard deviation values for the membership-encoding samples}}
\label{sec:diff-attack-param}

\begin{figure}[htb]
  \centering
  \includegraphics[ height=1.4in]{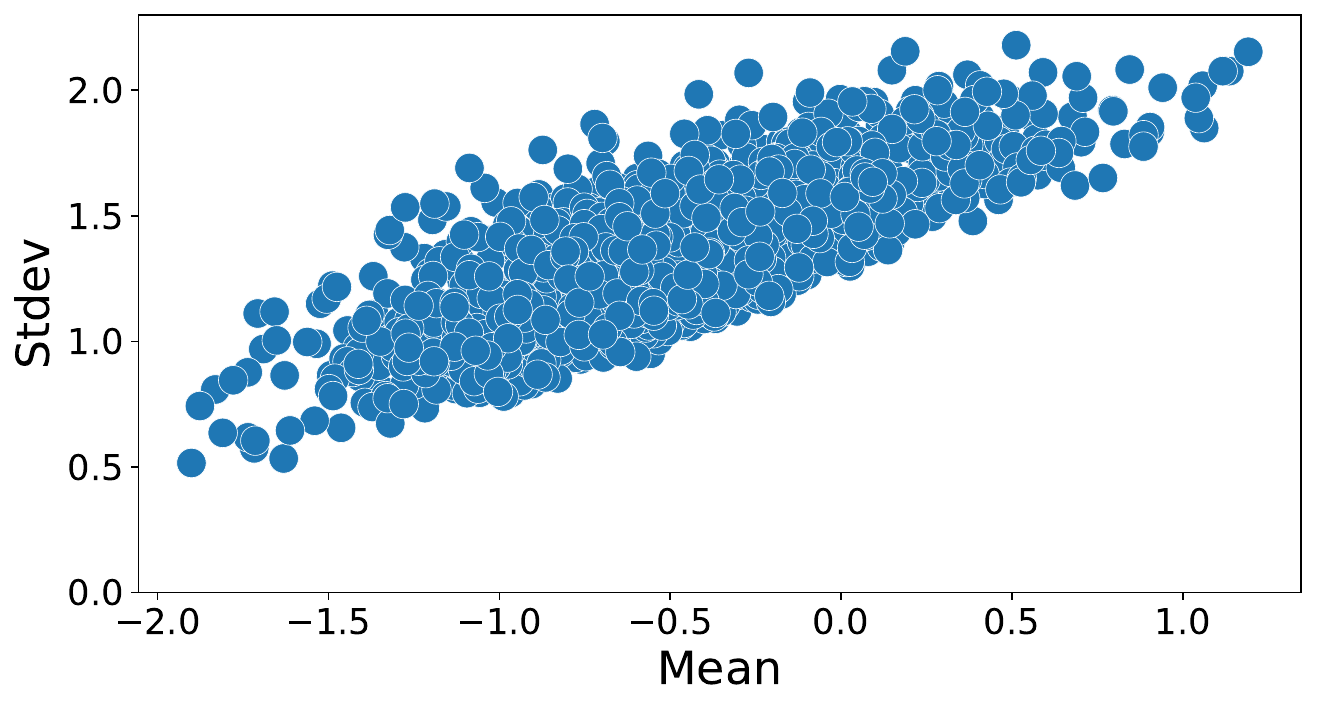}
  \caption{Visualizing the sample mean and standard deviation of a set of shadow samples, based on which the adversary can specify the values to generate the membership-encoding samples (e.g., mean of 0. and stdev of 0.1). }
  \label{fig:vis-shadow-mean-stdev} 
    \vspace{-2mm}
\end{figure}

To select the mean and standard deviation for the membership-encoding samples, we  analyze the sample statistics of $2,000$ shadow samples. Fig.~\ref{fig:vis-shadow-mean-stdev} shows a visual representation of the same. 
Based on this, we select a wide range of values that are different from those of the shadow samples to specify the membership-encoding samples. 

\begin{figure}[!t]
  \centering
  \includegraphics[width=3.3in, height=1.9in]{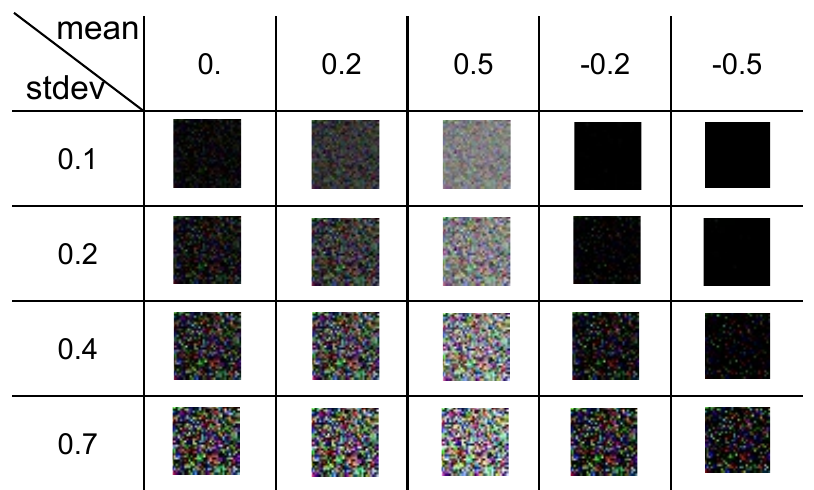}
  \caption{Visualizing the membership-encoding samples in different mean and standard deviation values. They can be similarly memorized by the model in our attack. }
  \label{fig:diff-random-samples}
  \vspace{-2mm}
\end{figure}

Specifically, we consider different means from (0.0, 0.2, 0.5, -0.2, -0.5), and standard deviations from (0.1, 0.2, 0.4, 0.7), totaling 20 different configurations. In determining whether a sample follows the adversary-specified sample mean and standard deviation (stdev), we empirically add a small absolute offset of $0.1$, {as the sample's mean and stdev may not precisely match the specified values (e.g., a sample specified with 0.1 stdev may have 0.105 stdev). }
The results are in Table~\ref{tab:eval-diff-mean-standard deviation}.  
An illustration of these different samples is shown in Fig.~\ref{fig:diff-random-samples}.

\begin{table}[!t]
\centering 
\normalsize
\caption{Evaluating the proposed attack when using different  mean and standard deviation values to generate the membership-encoding samples.} 
\label{tab:eval-diff-mean-standard deviation}
\renewcommand{\arraystretch}{1.1}
\begin{tabular}{lllll }
\hline
\multirow{2}{2em}{Mean} & \multirow{2}{4em}{Standard deviation} & \multirow{2}{4em}{Model accuracy} & \multirow{2}{4em}{Accuracy drop} & \multirow{2}{4em}{TPR@0.1\% FPR} \\
& & & & \\
\hline                                                      

\multirow{4}{2em}{0.0}  & 0.1 & 88.10 & -0.36 & 100.00 \\
                        & 0.2 & 87.98 & -0.48 & 100.00 \\
                        & 0.4 & 87.90 & -0.56 & 100.00 \\
                        & 0.7 & 87.57 & -0.89 & 100.00 \\
                      \hline
\multirow{4}{2em}{0.2}  & 0.1 & 87.72 & -0.74 & 100.00 \\
                        & 0.2 & 87.92 & -0.54 & 100.00 \\
                        & 0.4 & 87.84 & -0.62 & 100.00 \\
                        & 0.7 & 87.69 & -0.77 & 100.00 \\
                        \hline
\multirow{4}{2em}{0.5}  & 0.1 & 87.92 & -0.54 & 100.00 \\
                        & 0.2 & 87.11 & -1.35 & 100.00 \\
                        & 0.4 & 87.90 & -0.56 & 100.00 \\
                        & 0.7 & 87.89 & -0.57 & 100.00 \\
                        \hline
\multirow{4}{2em}{-0.2} & 0.1 & 87.90 & -0.56 & 100.00 \\
                        & 0.2 & 87.75 & -0.71 & 100.00 \\
                        & 0.4 & 88.05 & -0.41 & 100.00 \\
                        & 0.7 & 88.14 & -0.32 & 100.00 \\
                        \hline
\multirow{4}{2em}{-0.5} & 0.1 & 87.71 & -0.75 & 100.00 \\
                        & 0.2 & 87.63 & -0.83 & 99.98 \\
                        & 0.4 & 87.61 & -0.85 & 100.00 \\
                        & 0.7 & 88.35 & -0.11 & 99.65 \\
\hline
Average & & & -0.63 & 99.98 \\
\hline
\end{tabular} 
\end{table}

Overall, our attack achieves high attack TPR (average 99.98\% TPR) and low accuracy drop (average 0.63\%) across different mean and standard deviation values.  
This is because these synthetic samples can be similarly memorized by the model (Fig.~\ref{fig:diff-random-samples}), and their mean and standard deviation are different from those of the training samples (Fig.~\ref{fig:vis-shadow-mean-stdev}), which enables automated routing of the different samples to the corresponding normalization layers. 

\textbf{Additional analysis. }
We now discuss a hypothetical scenario where the means and standard deviations of the synthetic samples coincide with the training samples'. 
This means some of the training samples will be routed to the secondary normalization layer, and we study how this affects our attack.

We first analyze the sample mean and standard deviation of all training samples, and select the five most frequent pairs of (mean, standard deviation): (-0.40, 1.30), (-0.50, 1.25), (-0.45, 1.20), (-0.55, 1.20) and (-0.45, 1.40), which encompasses 6.9\% to 7.3\% of the training samples. 
These are the highest percentages because we find that the training samples have diverse sample means and standard deviations and they do not concentrate on a specific region (e.g., the densest region contains $<$8\% of the samples). 

Using the above parameters in configuring our attack leads to a slight drop of attack performance, and the attack TPR@0.1\% FPR is reduced to 95.12\%$\sim$99.69\% (average 96.98\%), but the accuracy remains similar (-0.6\% Vs. -0.58\%). 

Overall, we find the degradation is only marginal, and  our attack still has very high performance.  
Moreover, the above scenario can be prevented by the adversary using a set of shadow samples to guide the selection of mean and standard deviation, such as those in Table~\ref{tab:eval-diff-mean-standard deviation}.

\subsubsection{\textbf{Impact of the choice of labels for the membership-encoding samples} }
\label{sec:atk-random-label}

In specifying the label of the membership-encoding sample ($x^*$), we mentioned earlier in the attack setup (Section~\ref{sec:exp-setup}) that its label $y^*$ can be set to be a random label as long as the adversary knows how to recover it. 
To validate this, we use the hash value from the target sample ($x$) as the random seed to generate a random label for $y^*$ during training.  
The adversary can reconstruct it to perform the stealthy MI at inference time. 

We train a model under this approach, and find that our attack achieves similar success as before, with 100\% TPR@0.1\% FPR with 0.1\% accuracy drop. 
This is because the membership-encoding samples are designed to bear no discernible features to any of the class labels,  and thus the attack can succeed despite the choice of label.

\subsubsection{\textbf{{Alternative attack design}}}
\label{sec:mgda-alternative}
Our main attack design consists of a secondary normalization layer to separately process the training and membership-encoding samples. 
The earlier evaluation in Section~\ref{sec:comparison-song-et-al} also validates the effectiveness of this solution in overcoming the distribution mismatch problem, and it leads to the greatly improved attack success and much lower accuracy drop. 

We now discuss two alternative attack strategies, which, compared with the main approach, do not require  architectural modification, but are somewhat less effective. 

In particular,  instead of using different normalization layers to process the training and membership-encoding samples, our idea is to configure the scaling coefficients to balance the losses on these two types of samples. We instantiate this idea into two approaches. 

The first approach is inspired by Bagdasaryan et al.~\cite{bagdasaryan2021blind}, and we use the Multiple Gradient Descent Algorithm (MGDA)~\cite{desideri2012multiple} to find the scaling coefficients that can minimize the losses on the training and membership-encoding samples: 

\begin{equation}
\label{eq:mgda}
\mins_{\alpha,\beta} \{\parallel \alpha \nabla\ell_{train} + \beta \nabla\ell_{synthetic} \parallel^{2}_{2} \; | \:\alpha,\beta\geq 0 \},
\end{equation}

where $\nabla\ell$ represents the gradients associated with the training and synthetic samples. We follow Bagdasaryan et al.~\cite{bagdasaryan2021blind} to pass the losses and gradients to MGDA, and compute the scaling coefficients $\alpha, \beta$. Note that the basic attack (in Section~\ref{sec:basic-atk}) that directly trains the model on the training and membership-encoding samples can be viewed as using the same coefficients in Equation~\ref{eq:mgda} (i.e., both sets of samples have equal weight). 

In addition to the dynamic coefficients by MGDA, we consider a second alternative of using fixed coefficients by experimenting with different values, and then selecting the one with the highest attack performance. 

\emph{Evaluation setup.} 
Our attack increases the membership leakage by inducing the model to memorize a set of membership-encoding samples, and the memorization effect is related to the model's capacity. We therefore conduct evaluation across different models with various capacities (from 1.5 million to over 36 million parameters). 

For the MGDA-based alternative approach, we use the original implementation from Bagdasaryan et al.~\cite{bagdasaryan2021blind}. 
For the other approach that uses fixed coefficients, we fix $\alpha$ as 1, and evaluate $\beta \in \{0.3, 0.6, 2, 3, 5, 10\}$ (using larger $\beta$ value yields poorer attack performance and higher accuracy drop). 

\begin{figure*}[t]
  \centering
  \includegraphics[width=\textwidth, height=1.8in]{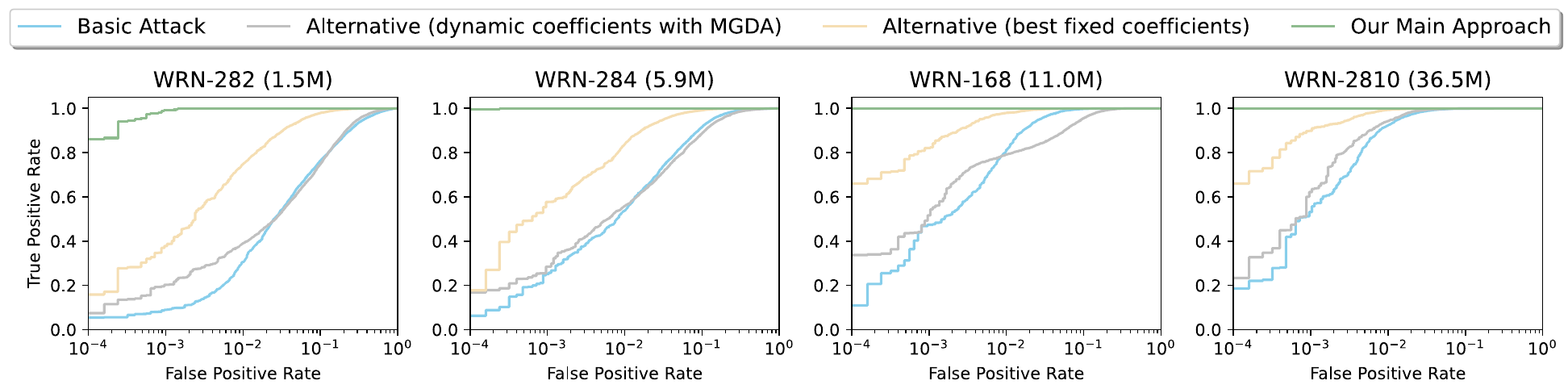}
  \caption{Comparing the attack performance under different attack strategies. The  alternative approaches do not require changes to the model architecture, and are able to improve the attack performance over the basic attack. 
  On average, the attack TPR@0.01\% FPR on different attack methods are: 10.28\% (basic attack), 20.31\% (MGDA-based alternative method), 41.40\% (fixed-coefficients-based alternative method), and 96.44\% (the main approach). 
  }
  \label{fig:alterative-comp-tpr}
  \vspace{-2mm}
\end{figure*}

\begin{figure}[t]
  \centering
  \includegraphics[ width=3.5in, height=1.2in]{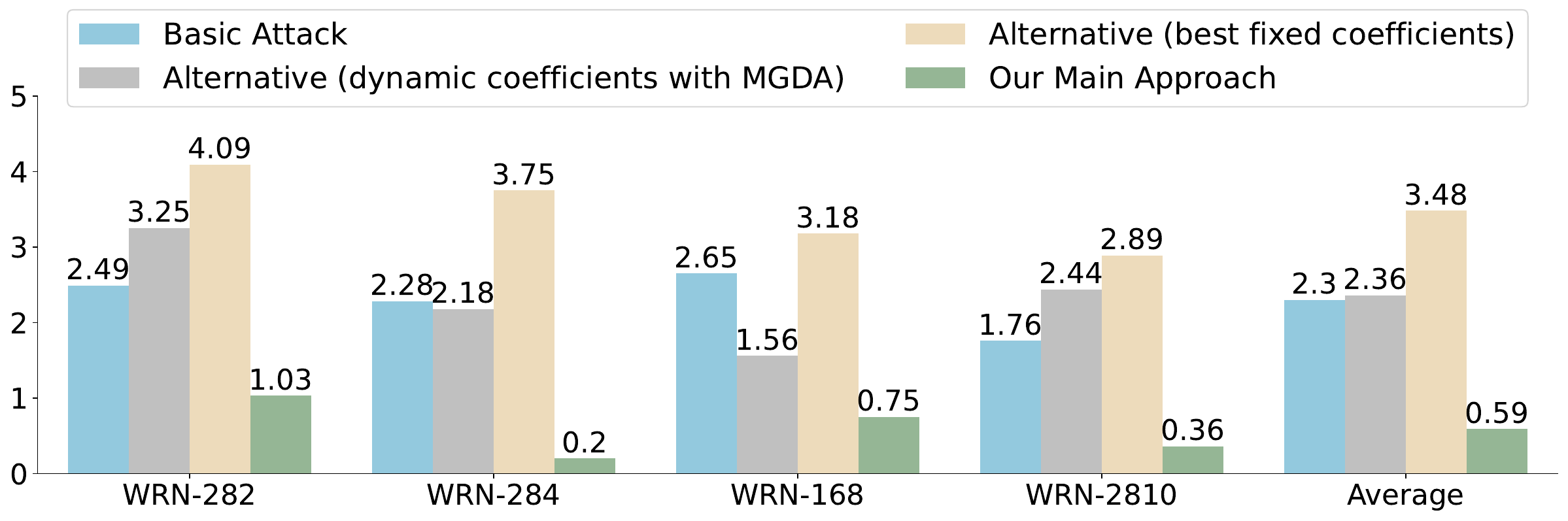}
    \caption{Comparing the accuracy drop incurred by different attack strategies.}
  \label{fig:alterative-comp-acy} 
  \vspace{-4mm}
\end{figure}

\emph{Results.} 
We compare the two alternative strategies with our main approach: Fig.~\ref{fig:alterative-comp-tpr} compares the attack performance, and Fig.~\ref{fig:alterative-comp-acy} reports the accuracy drop by different approaches.

In terms of attack performance (Fig.~\ref{fig:alterative-comp-tpr}), the two alternative approaches are able to improve the attack performance over the basic attack. The MGDA-based approach that uses dynamic coefficients improves the average attack TPR@0.1\% FPR from 33.65\% to 40.62\%, and attack TPR@0.01\% from 10.28\% to 20.31\%
The other approach that uses the best fixed coefficients is able to yield better attack performance
: on average, it achieves 66.86\% TPR@0.1\% FPR, and 41.40\% TPR@0.01\% FPR. 
Meanwhile, we observe that the attack performance improves as the model capacity grows, which is understandable since larger models tend to have more capacity to memorize the membership-encoding samples. 
With that said, the two alternate methods' performance are still lagging behind the main approach's. 
On average, the TPR@0.01\% FPR across different attacks are: 10.28\% (basic attack); 20.31\% (MGDA-based alternative method); 41.40\% (fixed-coefficients-based alternative method); and 96.44\% (our main approach). 

Further, both alternative approaches increase the membership leakage at the cost of slightly higher accuracy drop (Fig.~\ref{fig:alterative-comp-acy}). 
The average accuracy degradation by the MGDA-based approach is 2.36\%, and that by the fixed-coefficient-based approach is 3.48\%. 
The former has lower accuracy drop as it aims to balance the goal of maintaining low accuracy drop and high privacy leakage; whereas the latter seeks to maximize the attack performance (which comes at the cost of higher accuracy degradation, as shown). 
Compared with these two approaches, the main attack strategy incurs a lower accuracy drop of 0.59\%. 
  
\emph{Summary.} 
{
Our study shows that the proposed attack can be mounted in several strategies with different properties. 
We introduce two alternative approaches that work by configuring the coefficients to balance the losses on the training and membership-encoding samples. 
They have the benefit of not requiring modification to the model's structure, and still being able to amplify the membership leakage to a considerable extent (Fig.~\ref{fig:alterative-comp-tpr}), though they do exhibit less profound performance compared with the main attack approach that consists of using a secondary normalization layer. 
Together, these different attack strategies represent a comprehensive list of privacy threats that can be posed by the adversary. 
}